\documentclass[%
 reprint,
superscriptaddress,
 amsmath,amssymb,
 aps,spanish
]{revtex4-2}

\usepackage{graphicx}
\usepackage{dcolumn}
\usepackage{bm}
\usepackage{subcaption}
\usepackage{amsmath}
\usepackage{xfrac}
\usepackage{xcolor}
\graphicspath{ {./figs/} }

\begin{document}

\preprint{APS/123-QED}

\title{Scheme for sub-shot-noise transmission measurement\\
using a time multiplexed single photon-source}

\author{Agustina G. Magnoni}
\email{amagnoni@citedef.gob.ar}

\affiliation{%
 Laboratorio de Óptica Cuántica, DEILAP, UNIDEF (CITEDEF-CONICET), Buenos Aires, Argentina}%
 \affiliation{Departamento de Física, Facultad de Ciencias Exactas y Naturales, UBA, Ciudad de Buenos Aires, Argentina}
\author{Laura T. Knoll}%
\affiliation{INRIM Istituto Nazionale di Ricerca Metrologica, Turin, Italy}
\affiliation{%
 Laboratorio de Óptica Cuántica, DEILAP, UNIDEF (CITEDEF-CONICET), Buenos Aires, Argentina}%
 \author{Miguel A. Larotonda}%
\email{mlarotonda@citedef.gob.ar}
\affiliation{%
 Laboratorio de Óptica Cuántica, DEILAP, UNIDEF (CITEDEF-CONICET), Buenos Aires, Argentina}%
 \affiliation{Departamento de Física, Facultad de Ciencias Exactas y Naturales, UBA, Ciudad de Buenos Aires, Argentina}




\date{\today}

\begin{abstract}
A promising result from optical quantum metrology is the ability to achieve sub-shot-noise performance in transmission or absorption measurements. This is due to the significantly lower uncertainty in light intensity of quantum beams with respect to their classical counterparts. In this work, we simulate the outcome of an experiment that uses a multiplexed single-photon source based on pair generation by continuous spontaneous parametric down conversion (SPDC) followed by a time multiplexing set-up with a binary temporal division strategy \cite{schmiegelow2014multiplexing}, considering several types of experimental losses.
With such source, the sub-Poissonian statistics of the output signal is the key for achieving sub-shot-noise performance. We compare the numerical results with two paradigmatic limits: the \textit{shot-noise limit} (achieved using coherent sources) and the quantum limit (obtained with an ideal photon-number Fock state as the input source). We also investigate conditions in which threshold detectors can be used, and the effect of input light fluctuations on the measurement error. Results show that sub-shot-noise performance can be achieved, even without using number-resolving detectors, with improvement factors that range from 1.5 to 2. This technique would allow measurements of optical absorption of a sample with reasonable uncertainty using ultra-low light intensity and minimum disruption of biological or other fragile specimens.

\end{abstract}

\maketitle


\section{\label{sec:intro} Introduction}

The field of quantum metrology and its applications in biological sciences \cite{giovanetti2011,taylor2016quantum,wasielewski2020exploiting,magana2019quantum,pang2017optimal} is a hot research topic. In particular, much attention has been paid to the utilization of quantum light as a resource for surpassing the classical limit of precision per unit intensity \cite{berchera2019quantum,moreau2019imaging}. The ability to obtain a light source that operates below the shot-noise limit - for a given limited level of probe beam intensity - allows for performance improvement of tasks such as imaging of photo-reactive biological samples \cite{meda2017photon}, high precision optical activity measurements in chiral media \cite{yoon2020experimental}, and high sensitivity single biomolecule detection and tracking \cite{mauranyapin2017evanescent} among others.

This work is devoted to study the behavior of an engineered light source that emits non-classical light states in the task of measuring the transmittance of a sample, in order to obtain an enhancement on the precision when compared to a measurement using classical light. Generally speaking, the lower the intrinsic uncertainty in the intensity of the incident light, the better the result will be. The use of low intensity light sources with sub-Poissonian statistics for absorption and transmission measurements is a promising technique for the study of fragile biological samples and ultra-sensitive materials, since experiments can be performed with minimum disturbance \cite{zonios2008melanin,cone2015measuring,mauranyapin2017evanescent}.

The uncertainty of such measurement is given by the combination of random fluctuations inherent in the optical probe beam, and the stochastic nature of the interaction between light and matter within the sampled object. By modeling two kinds of light sources, such as coherent states and Fock or number states, different precision limits can be obtained. In the former case, using a light source with Poissonian photon statistics to measure the transmittance of an object leads to a precision in the measurement that is bounded by the \emph{Shot-Noise Limit} (SNL). Instead, the eventual use of an ideal antibunched light source such as a Fock state photon source gives the \emph{Ultimate Quantum Limit} (UQL) for the measurement precision, due to the deterministic nature of the photon number emission. A true Fock state photon source, however, is not yet an available resource. The challenge is therefore to obtain an engineered light source with intensity fluctuations below the Poissonian limit and to combine it with an adequate choice of an estimator, in order to obtain a measurement scheme that outperforms the classical one.

Real-world single-photon sources are an interesting option since an ideal $N$-photon Fock state as input achieves the same precision as an ideal single photon input and $N$ repetitions. High photon-number Fock states are not experimentally achievable nowadays, but a great deal of research is currently in progress to obtain devices that deliver light pulses carrying a single photon in well-defined spatio-temporal and polarization modes. 
Different approaches rely on either the use of single emitter sources, or some kind of multiplexing of one or several heralded photon pair sources based on  Spontaneous Parametric Down Conversion (SPDC). Single emitter sources include devices based
on fluorescence from atoms, ions or molecules  \cite{diedrich1987nonclassical,kimble1977photon,mandel1979sub,basche1992photon} and on different types of ``artificial atoms''. Single photon emission from Nitrogen-vacancy centers has been extensively reviewed in \cite{aharonovich2016solid}.  Quantum dots have been known to emit single photons since the end of the last century \cite{couteau2004correlated,holmes2014room,sebald2002single,gammon1996homogeneous,michler2000quantum}. By coupling the quantum dots with electrically controlled cavities in deterministically fabricated devices, an enhancement on brightness and purity of these kind of sources has been obtained, reaching indistinguishabilities above 99\% and photon extraction efficiencies on the order of 66\% \cite{QDpan,somaschi2016near}. Recently, the effect of imperfections and unwanted multi-photon components of such sources on the quality of the source has been studied both theoretically and experimentally \cite{ollivier2020hong}. 

Another approach to the single photon source is based on nonlinear processes such as SPDC or four-wave mixing, where a pair
of photons is produced after the interaction of one or two pump photons with a nonlinear material. One of the output photons is
sent to a detector, which heralds with high probability the presence of the other photon, provided the two
downconverted modes are non-degenerated in some degree of freedom. State-of-the-art implementations involve integrated optics devices and heralding efficiencies exceeding 50\% \cite{bock2016highly,montaut2017high} and even reaching 90\% \cite{paesani2020near}. The heralding process removes the zero photon component of the heralded field, but there is still a non-negligible probability of generating more than one pair, that scales with the pump intensity. Moreover, since these processes are probabilistic, there is a trade-off between the probability of generating a photon and the fidelity of the output to a single-photon state. By spatially multiplexing several heralded sources \cite{migdall2002tailoring,shapiro2007demand,jennewein2011single,ma2011experimental,christ2012limits,collins2013integrated,mazzarella2013asymmetric}, temporally multiplexing a single source \cite{jeffrey2004towards,mower2011efficient,glebov2013deterministic,schmiegelow2014multiplexing,kaneda2015time,mendoza2015active,rohde2015multiplexed,zhang2017indistinguishable,kaneda2019high}, or even combining temporal and spatial multiplexing \cite{adam2014optimization}, the source brightness can be (ideally) arbitrarily increased, depending on the size of the multiplexing network and its overall throughput.    
A comprehensive and updated review on these kind of sources can be found in \cite{meyer2020single}.
 
Meanwhile, transmission/absorption measurements have been widely studied in recent years. There are several different approaches to the task of achieving the highest possible precision, that rely on different estimators and light sources, with and without spatial resolution. Schemes for estimating the transmission of a sample generally consist in measuring the intensity attenuation of a light beam traversing it, which can be done with single beam light sources or with twin-beam correlated sources. Twin-beams and difference-based estimators have been used for spatially resolved implementations \cite{brida2010experimental} including the realization of the first sub-shot-noise wide field microscope in 2017 \cite{samantaray2017realization}. The performance of this estimator depends upon the spatial resolution and reaches out a factor of improvement in precision over the SNL of approximately 1.30. Estimators based on the ratio of two correlated beams have been first proposed in \cite{jakeman1986use} and its recent experimental implementations use heralded single-photon sources and achieve a maximum improvement factor of 1.79 \cite{sabines2017sub,sabines2019twin}. Another estimator based on the ratio of two signals but with some optimizations is presented in \cite{moreau2017demonstrating}, reaching a maximum improvement factor of 1.46. Recently, a complete theoretical and experimental study of the performance of these different estimators was presented \cite{losero2018unbiased}. 

In this work we propose to use the correlations present in a photon pair SPDC source to build a specific time multiplexed single photon source \cite{schmiegelow2014multiplexing,magnoni2019performance} and use it as input in a direct-type measurement of the transmission. We compare its performance both with a coherent light source (with Poissonian statistics, i.e. a ``classical'' experiment) and a perfect single-photon Fock state. The results of this work can be adapted in a straightforward manner to essentially any temporally multiplexed source that uses either fiber loops, ring cavities or fixed length delays to re-route the heralded photons.  

Another issue to take into account in order to maximize measurement precision is the performance of the detection devices. Photon number-resolving detectors are the ultimate refinement for intensity detection and constitute the most sophisticated measurement devices for quantum optics. Different technological approaches are currently employed to obtain high efficiency detectors with photon number resolution, such as arrays of multiplexed standard detectors  \cite{achilles2003fiber,fitch2003photon}, Transition Edge Sensors (TES) \cite{lita2008counting,fukuda2011titanium,gerrits2016superconducting}, Superconducting Nanowire Photon Detectors (SNPD) \cite{natarajan2012superconducting,marsili2013detecting,cahall2017multi}, and even Coupled Charged Devices (CCDs) with floating-gate amplifiers \cite{tiffenberg2017single}. However, photon counting detectors with low dark counts and mid-to-high efficiency such as avalanche photodiodes that are unable to resolve the number of detected photons (threshold detectors), are quite widespread and their use is very common in research and metrology laboratories. We therefore study the behavior of the proposed set-up under two different detection schemes: number resolving and threshold detectors. 

This paper is organized as follows: in section \ref{sec:fuente} we briefly review the main features of  the proposed single-photon source. Section \ref{sec:estimators} is devoted to the discussion on the different transmission estimators used under the aforementioned detection conditions, and the discussion on the expected advantage that can be obtained with the single-photon source. In section \ref{sec:fluctuations} we analyse the effect of fluctuations on the transmission estimation when the sources are fed with a photon pair source with super-Poissonian photon statistics. Final remarks and comments regarding the prospective for the use of single-photon sources to obtain quantum advantage on transmission measurements are pointed out in section \ref{sec:final}.

\section{\label{sec:fuente} Binary-time Multiplexed Single-Photon source}

The sub-Poissonian light source model studied herein is named Binary-time Multiplexed Single-Photon (BinMux-SP) source. This device is based on photon pair generation by SPDC and a time multiplexing stage that both raises the single-photon probability and synchronizes the output state to an external clock signal.

\begin{figure}[h]
   \includegraphics[width=0.45\textwidth]{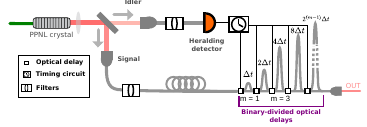}
   \caption{The Binary-time Multiplexed Single-Photon (BinMux-SP) source consists of a collinear, non-degenerate photon pair generation stage, and a time multiplexing stage that raises the single-photon probability and synchronizes the output photon to an external clock signal based on the timing of the herald photon detection. The latter stage is based on an array of several fiber optic fixed length paths, each one imposing a different temporal delay to the signal photon.}
   \label{fig:BinMux}
 \end{figure}

It specifically consists of a single heralded photon source and a network of fiber optic components that conform the time multiplexing stage (Fig. \ref{fig:BinMux}). This array has several possible fiber paths with different lengths, that consequently impose different temporal delays to the signal photon. Which-path decision is made by a timing circuit, based on the information given by the detection of the idler photon. Given the external clock signal with period $T$, the time multiplexing stage locates the signal photon (within a fixed short temporal window $\Delta t$) and re-routes it to the output, synchronized to the clock tick.  

The time multiplexing network is binary-divided with $m$ possible delay stages, each one with a duration of $(2^{(i-1)} \Delta t)$, $i\in \{1,\cdots,m\}$. Active switching of this stages allows for compensation of a temporal mismatch (between the photon and the clock) up to $\mathcal{T}=(2^m-1)\Delta t$. A detailed description of the source can be found in \cite{schmiegelow2014multiplexing}, and the effect of non-ideal components such as optical losses, detector dark counts and detector efficiency was presented in \cite{magnoni2019performance}. 

The minimum correcting time $\Delta t$ is set by the coherence length of the photon pair and eventually by the photo-detection jitter at the heralding side. The efficacy of the setup as a sub-Poissonian photon gun
relies on the ability to obtain a high probability of detecting at least one photon during the total synchronization interval $T$ - thus minimizing the zero-photon component of the distribution- together with a low probability of multi-photon occurrence within a single temporal window $\Delta t$. Optimum conditions for fixed number of delay stages (or \textit{correction stages}) and loss can be obtained by adjusting the input mean photon pair number $\mu$ from the SPDC source.

\section{\label{sec:estimators} Transmission estimators}

\begin{figure}[h]
   \includegraphics[width=0.45\textwidth]{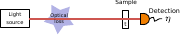}
   \caption{General scheme for a direct transmission measurement: a light source sends a beam through the sample and the transmitted light is detected afterwards. The detector efficiency is $\eta = 0.9$ and it may or may not have number-resolving capacity. Lenses and other optical elements can impose an additional loss between the source and the sample. The source studied in this work is the proposed Binary-time Multiplexed Single-Photon source, which is also benchmarked against a coherent source with Poissonian photon statistics and a perfect single-photon Fock state. For the non-ideal sources we consider an optical loss of $0.9$, while we do not consider any optical loss for the ideal Fock-state photon source.}
   \label{fig:setup}
 \end{figure}

Following the strategy of using many single photons instead of a large number Fock state as input, we study the performance of the BinMux-SP in a direct single beam transmission measurement scheme, like the one depicted in Fig.~\ref{fig:setup}. We compare its performance by replacing the BinMux-SP with a coherent state source and with a perfect single-photon Fock state. We consider an optical loss of $0.9$ for the coherent and BinMux-SP cases, while we use a lossless channel for the Fock state source. The efficiency of the detectors ($\eta$) is, in all cases, considered to be always $0.9$. 

The performance of the different estimators ($\hat{T}$) of a given parameter ($t$) can be compared by computing their expected value E, and their mean squared error (MSE). These quantities depend on the random variable $k$ that is being measured and its probability distribution $P(k)$ \cite{hayat1999reduction,mood}:
\begin{align}
\label{ec:mean}
    & \text{E}(\hat{T}) = \sum_k \hat{T}(k) P(k), \\
\label{ec:MSE}
    & \text{MSE}(\hat{T}) = \sum_k [\hat{T}(k) - t]^2 P(k),
\end{align}
for discrete probability distributions. Throughout this work, $k$ represents the number of detected events on each experiment. The benefit of using the MSE over the variance ($\text{Var}(\hat{T}) = \sum_k [\hat{T}(k) - \text{E}(\hat{T})]^2 P(k)$), is that it takes into account both the precision and the accuracy of the estimator. This quantities enable us to simulate the performance of the estimators using the real value of the parameter. 

In what follows we introduce transmission estimators both for number-resolving (NR) and for threshold detectors. Throughout the work we assign the SNL performance to the coherent state photon source illuminating a translucent sample using a number-resolving detector, and the UQL performance with the single-photon Fock state which ---given its nature--- does not change with the choice of the detector.  

\subsection{\label{subsec:NR} Estimators for number-resolving detectors}

When using number-resolving detectors, the measured random variable is the number of photons. For the three different sources considered in this work, the probability distributions are: Poissonian in the coherent case, sub-Poissonian in BinMux-SP \cite{magnoni2019performance} and a perfect number state -with null variance- in the single-photon Fock case.

The general form of the transmission estimators for each case are:
\begin{align}
      &\label{ec:Tc} \text{Coherent:} \hspace{9mm} \hat{T}^{nr}_c(k_c) = \frac{k_c}{\eta * \langle n_c \rangle}. \\[5pt] 
      &\label{ec:Tb} \text{BinMux-SP:} \hspace{5mm} \hat{T}^{nr}_b(k_b) = \frac{k_b}{\eta * \langle n_b \rangle}. \\[5pt]
      &\label{ec:Tf} \text{Fock:} \hspace{16mm} \hat{T}^{nr}_f(k_f) = \frac{k_f}{\eta N_{in}}. 
\end{align}

The measured number of photons in the experiments with coherent, BinMux-SP and Fock sources is $k_c$, $k_b$ and $k_f$ respectively; $\eta$ is the detector efficiency (which is set to $0.9$ throughout the simulations), $\langle n \rangle$ is the mean number of photons incident on the sample and $N_{in}$ is the eigenvalue of the photon number operator in the case of Fock states. In particular, we set $N_{in} = 1$ and $\langle n_{c} \rangle = \langle n_{b} \rangle = 1$.

Since the relative error is high in such a low intensity regime, we considered $\nu = 200$ number of repetitions of the experiment, in order to reduce the uncertainty of the measurement.  

\begin{figure}[h]
   \includegraphics[width=0.49\textwidth]{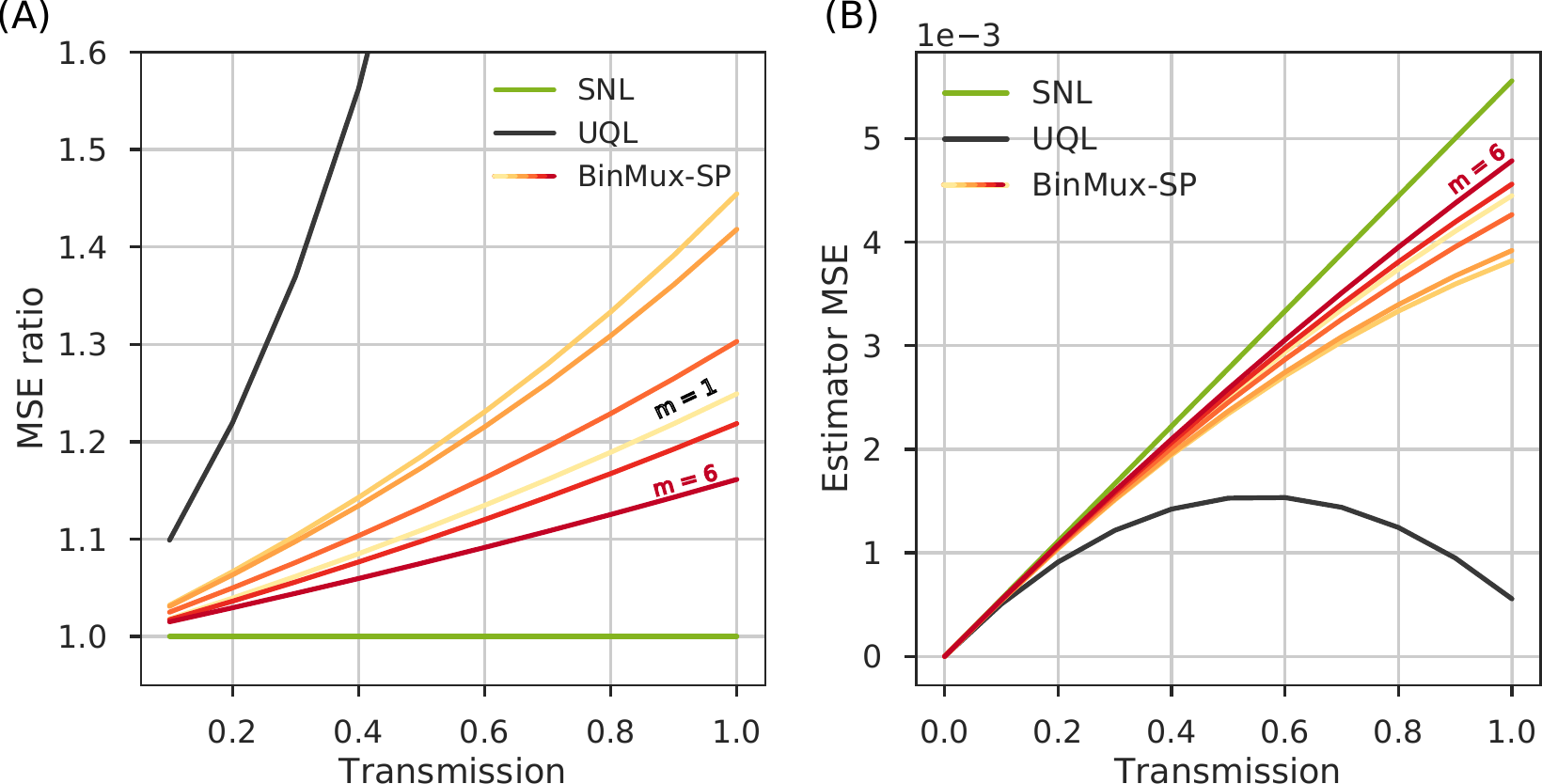}
   \caption{(A) MSE ratio between the coherent source (SNL) and the BinMux-SP (with 1 to 6 correction stages $m$) or the single-photon Fock state (UQL). (B) MSE for the SNL, the BinMux-SP source and the UQL. Both graphs correspond to $\nu = 200$ repetitions and number-resolving detectors. Input mean photon numbers are: $\langle n_{c} \rangle = \langle n_{b} \rangle = 1$.}
   \label{fig:ratioNR}
 \end{figure}

The first important property of these estimators is that they are accurate (or unbiased) for all three sources: the expected value is equal to the transmission parameter, which also means that the MSE is equal to the variance. In order to study the improvement factor, we compute the ratio between the MSE of the coherent source (what we call the SNL) to the MSE of each source [Fig.~\ref{fig:ratioNR} (A)]. For the BinMux-SP, we studied the performance of different amount of correction stages $m$ (sec.~\ref{sec:fuente}). Analyzing the ratios enables for a quick check of the quantum advantage (ratio $> 1$) and also independence from the number of repetitions ($\nu$). Repetitions only raise the precision equally for all sources (in the case of unbiased estimators), reducing the variance by a factor $\nu$. The MSE as a function of the transmission alone can be observed in Fig.~\ref{fig:ratioNR} (B), for $\nu = 200$.  

It is clear that for the case of the BinMux-SP source, there is an advantage over the shot-noise limit for the complete transmission range. Different amount of $m$ stages have different performances; since the loss introduced by the optical components eventually limits the performance of the source, the maximum ratio is obtained for $m = 2$. Regarding the absolute value of the MSE (or variance, in this case), it can be further reduced by increasing the number of repetitions $\nu$. It is important to note that these estimators approach the limit imposed by the Cramér-Rao bound for the variance \cite{frodesen}. Therefore, the estimator is close to optimum for the probability distribution of the BinMux-SP source.

\subsection{\label{subsec:theshold} Threshold detectors}

Even though there has been substantial attention and significant progress in the development of photon number-resolving detectors \cite{Cahall:17}, they are still expensive and resource-demanding pieces of equipment. On the other hand, standard photon counting devices with binary output (detection-no detection) are quite common in quantum optics and metrology laboratories.  It is therefore interesting to study the changes introduced by replacing the number-resolving detectors with threshold detectors, considering the high efficiencies achieved \cite{Reddy:19} and the low intensity working scenario.

By replacing the number-resolving detectors, the measured random variable changes from number of photons to detector clicks, each one of them triggered by the arrival of one o more photons. The previous estimators (eqs. \ref{ec:Tc} - \ref{ec:Tf}) can be adapted in a straightforward fashion: in this condition $k^{t}$ represents the total number of clicks after the sample and $ \langle n^t \rangle$ the mean number of clicks without sample, in $\nu$ repetitions of the experiment:    

\begin{align}
      &\label{ec:Tc_th} \text{Coherent:} \hspace{9mm} \hat{T}^{t}_c(k^t_c) = \frac{k^t_c}{\langle n^t_c \rangle}. \\[5pt] 
      &\label{ec:Tb_th}\text{BinMux-SP:} \hspace{5mm} \hat{T}^{t}_b(k^t_b) = \frac{k^t_b}{\langle n^t_b \rangle}. \\[5pt]
      &\label{ec:Tf_th}\text{Fock:} \hspace{16mm} \hat{T}^{t}_f(k^t_f) = \frac{k^t_f}{\eta N_{in}}. 
\end{align}

The probability distributions of $k^t$ are now binomials $B(k^t|p^{click},\nu)$, with $\nu$ trials, $k^t$ successes and a probability of success per trial $p^{click}$. These probabilities can be calculated as:

\begin{align}
      &\label{ec:pc} \text{Coherent:} \hspace{9mm} p^{click}_{c} = \sum_{i \geq 1} [1 - B(0| t\eta,i)]P(i). \\[5pt] 
      &\label{ec:pb}\text{BinMux-SP:} \hspace{5mm} p^{click}_{b} = \sum_{i \geq 1} [1 - B(0| t\eta,i)]P_{b}(i). \\[5pt]
      &\label{ec:pf}\text{Fock:} \hspace{16mm} p^{click}_{f} = t\eta. 
\end{align}

$P(i)$ is the Poisson probability mass function for $i$ photons and $P_b(i)$ is the probability mass function of emitting $i$ photons of the BinMux-SP \cite{magnoni2019performance}. For each source, these $p^{click}$ probabilities correspond to the chance of emitting $i$ photons and at least one of them being detected, taking into account losses, sample transmission and detector efficiency. Because of the nature of the detection process, an increase of the multi-photon emission probability of the source lowers the correlation between a single-photon and a click. 

\begin{figure}[h]
   \includegraphics[width=0.4\textwidth]{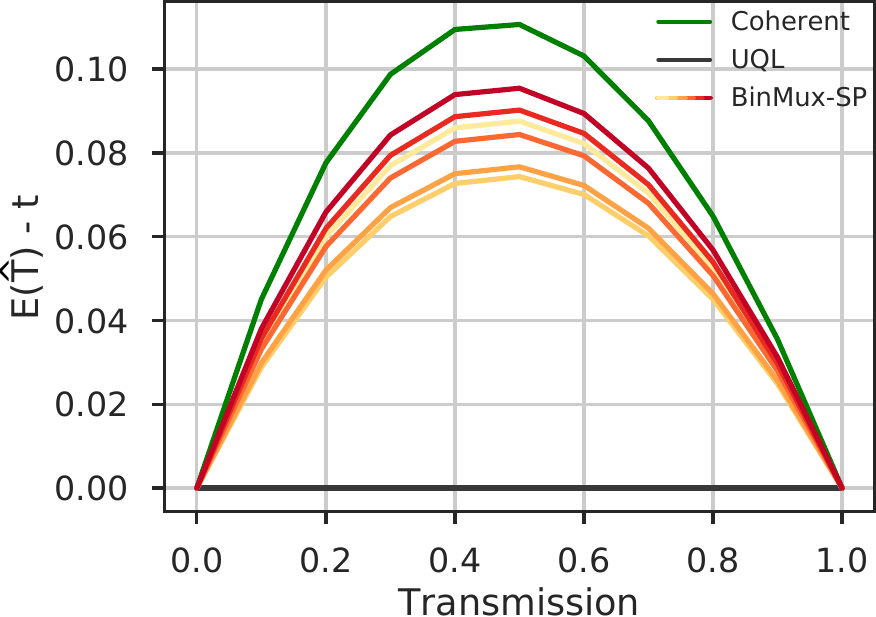}
   \caption{Difference between the mean of the estimator and the transmission as a function of the latter for the three sources (coherent, BinMux-SP and single-photon Fock states (UQL)) for $\nu = 200$ repetitions, using threshold detectors. Input mean photon numbers are: $\langle n_{c} \rangle = \langle n_{b} \rangle = 1$.}
   \label{fig:mean}
 \end{figure}

As expected, these estimators are biased for the coherent and the BinMux-SP cases, and unbiased for the Fock states. This can be seen in Fig. \ref{fig:mean}, which shows the difference between the mean of the estimator and the true value of the transmission. 

\begin{figure}[h]
   \includegraphics[width=0.47\textwidth]{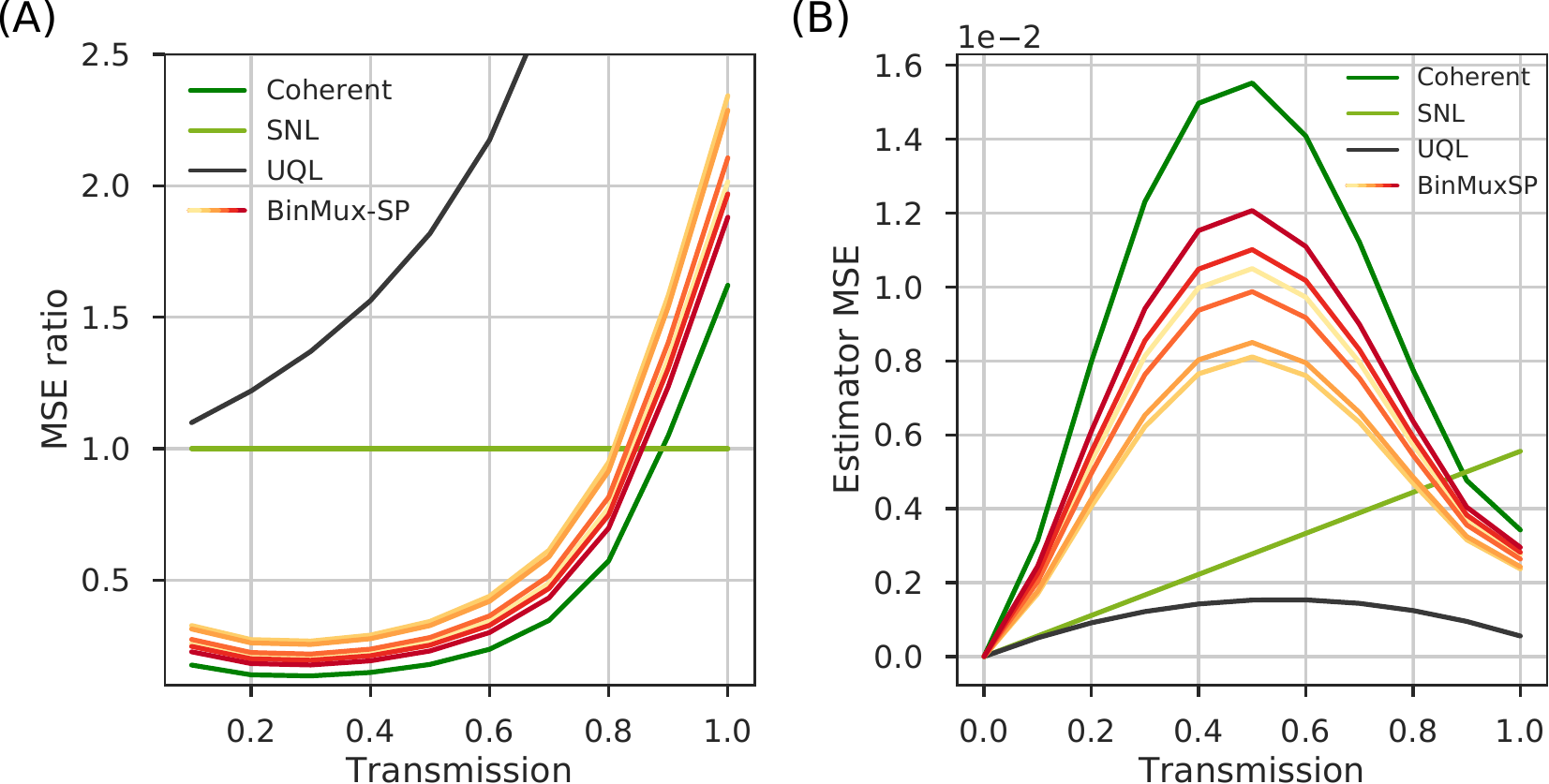}
   \caption{(A) MSE ratios for the threshold detection case. The reference curve (from which the ratios are computed) is again the one obtained using a coherent source with number-resolving detectors (SNL). The dark green curve corresponds to the coherent source, yellow-to-red curves for the BinMux-SP (using up to six correction stages $m$) and the black curve is the MSE ratio for a single-photon Fock state (UQL). (B) MSE for the same three sources, together with the SNL reference (light green curve) for direct comparison. Both plots correspond to $\nu = 200$ repetitions and input mean photon numbers $\langle n_{c} \rangle = \langle n_{b} \rangle = 1$.}
   \label{fig:ratioT}
 \end{figure}

The sub-Poissonian statistics of the BinMux-SP source are responsible for the reduced bias in the estimators. This behaviour can be observed in the MSE ratios and in the MSE alone shown in Fig. \ref{fig:ratioT} (A) and (B).
In this case, using the BinMux-SP source is favorable only for a limited range of transmissions close to the transparent $t=1$ limit. However, in this region the performance of the SNL can be significantly improved (in some cases with a ratio higher than 2). Additionally, it is always a better alternative to the coherent source with a threshold detector. It is also worth to note that for high transmissions, an advantage over the SNL can be obtained by using a threshold detector with the coherent state. This is due to the variance-reduction effect of the binarization introduced by the threshold detection, which has only two possible outcomes: click or no-click \cite{blanchet2008}.

However, unlike the case of unbiased estimators, it is not possible to arbitrarily reduce the MSE by increasing the number of repetitions $\nu$. The MSE has a fundamental limit, given by the inaccuracy of the estimators: it can not be smaller than the squared of the difference between the mean and the parameter (Fig. \ref{fig:mean}). However, since the estimator is not biased for $t=0$ and $t=1$, a good performance can always be obtained for high transmission values.   

Even though a fair comparison with the single-photon Fock state would correspond to input mean photon numbers $\langle n_{c} \rangle = \langle n_{b} \rangle = 1$, and given that the bias in the threshold estimators is mainly due to the multi-photon components of the statistics, it is interesting to explore the performance of the estimators for lower values of intensity. In the next section we study how these results are modified for different input intensities.

\subsection{Lower light intensity regimes}
\label{subsec:lowint}

\begin{figure}[h]
   \includegraphics[width=0.47\textwidth]{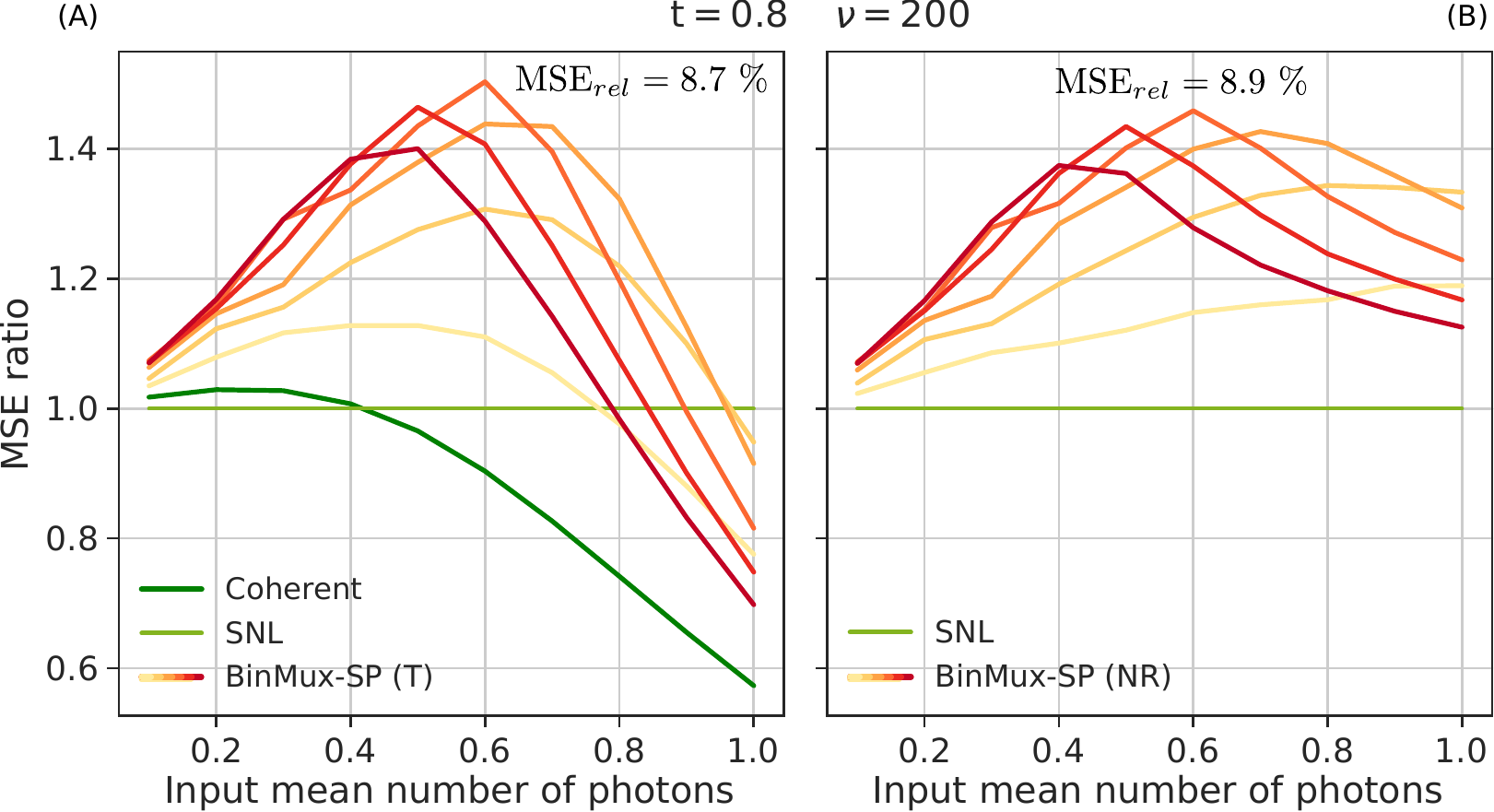}
   \caption{MSE ratio for a $t=0.8$ sample transmission as a function of the input mean number of photons ($\langle n \rangle$) for: (A) the threshold estimators and (B) the number-resolving estimators. These results are for $\nu = 200$. The relative MSE for the best performances of each estimator is added to the graph. The vertical axis is shared.}
   \label{fig:barrT}
 \end{figure}

Lowering the mean number of photons in order to reduce the amount of multi-photon emission is a common strategy when using a weak coherent pulse source. The sub-Poissonian nature of the BinMux-SP statistics also guarantees a more efficient response to this action, showing an improved reduction on the multi-photon pulses compared to that of the coherent source. The results obtained in the performance of the estimators by changing the mean photon number incident to the sample ($\langle n \rangle$) are shown in [Fig.~\ref{fig:barrT}] for a transmission $t=0.8$ and for $\nu = 200$ repetitions of the experiment.    

Although for a unity input mean photon number, the estimators with threshold detectors do not present any enhancement over the SNL (for $t = 0.8$), an advantage can be still obtained for lower intensities. Indeed, MSE ratios of $\sim 1.5$ can be achieved, which is very similar to the best performance available with number-resolving detectors. It is interesting to note that, for this transmission range and number of repetitions, the coherent source with threshold detection barely outperforms the SNL between 0 and $0.4$ mean photon number. This behavior is representative of the performance of the source for mid-range transmissions. This analysis enables a rapid visualization for selecting the most convenient combination of number of correcting stages $m$ on the BinMux-SP source and input mean photon number, for a given value of the transmission.

\begin{figure}[]
   \includegraphics[width=0.45\textwidth]{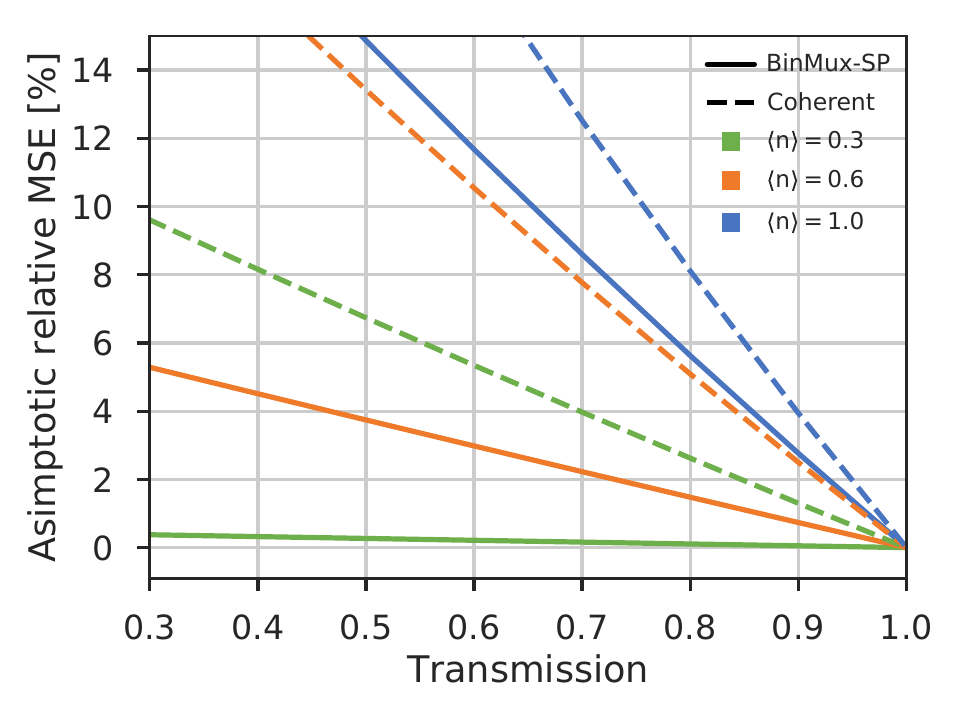}
   \caption{Asymptotic minimum relative MSE [\%] achievable as a function of the transmission. These values correspond to an infinite number of repetitions $\nu$ and are thus caused by the inaccuracy of the estimators. Solid line corresponds to the BinMux-SP source and the dashed line to the coherent source.}
   \label{fig:relative}
 \end{figure}

 Given that the estimators for threshold detectors are indeed biased, the improvement of their performance is therefore limited by the asymptotic limit imposed by their inaccuracy, while the number-resolving detector estimators only get more precise. In Fig. \ref{fig:relative} we present the asymptotic minimum relative MSE achievable per transmission value, for three representative values of input mean photon number, both for the coherent and the BinMux-SP sources.

 These values of MSE can be accessed by performing a very large number of repetitions; in this situation the inaccuracy eventually dominates over the precision. As expected, the MSE$\%$ is greatly influenced by the intensity used. The enhancement obtained with the use of the BinMux-SP source over a coherent source is present for all transmissions in all three values of $\langle n \rangle$. In this case, for mid-range transmission values a three-fold increase can be obtained using the BinMux-SP source instead of the coherent source.

\section{\label{sec:fluctuations} Effect of fluctuations in the input light source}

Both the BinMux-SP and the coherent sources must be fed with a light source governed by Poisson statistics; in the coherent case it is the source itself, whereas the BinMux-SP is based on a parametric fluorescence photon pair source, which for practical purposes, shows Poissonian statistics on the pair emission.
In order to test the robustness of the transmission measurement against excess fluctuations on the light source we consider the mean photon number $\mu$ as a random variable with Gaussian statistics. 

The mean value of this distribution is the one that fulfills the condition of light intensity incident on the sample (which for the following discussion is chosen to be $\langle n_c \rangle = \langle n_c \rangle = 0.5$), and the standard deviation corresponds to a certain percentage of the mean. We let this deviation vary from 0 to 60\% ($\sigma = a\mu$, $a\in \{0,\cdots,0.6\}$).

\begin{figure}[h]
   \includegraphics[width=0.47\textwidth]{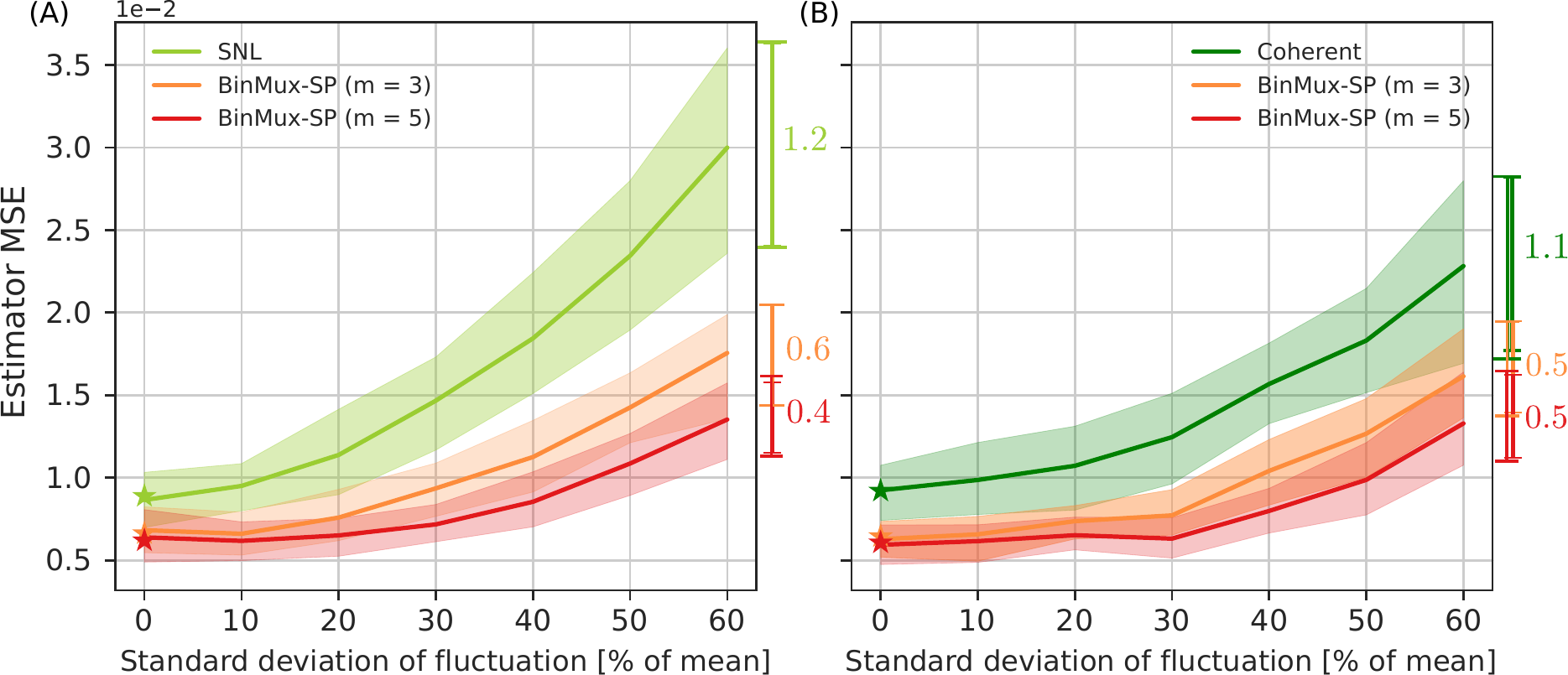}
   \caption{Estimator MSE for (A) the number-resolving detectors and (B) the threshold detectors, as a function of the amount of fluctuation present in the incident light (the standard deviation as a percentage of the mean). The solid line corresponds to the mean MSE (obtained over 50 rounds) whereas the shaded region shows the 68 \% confidence interval. The behavior with $m=3$ and $m=5$ correcting stages is shown. Figures at the right of each plot correspond to the confidence interval for the maximum fluctuation considered.}
   \label{fig:fluctuation}
 \end{figure}

The estimator MSE for a transmission $t = 0.8$ as a function of the size of the fluctuations (i.e. $a$ in \%) for both types of detectors is shown in Fig. \ref{fig:fluctuation}, for two representative amounts of correction stages $m$ for the BinMux-SP.

 Since the reference beam in this direct type of measurements is computed without fluctuations (or considering a significant amount of integration time that ensures compensation), the larger the fluctuations, the greater is the MSE observed; this is primarily caused by the bias introduced in the measurement. For the coherent source in the number-resolving case the MSE at maximum fluctuation is 3.4 times its original value, while for $m = 5$ BinMux-SP it is 2.1. This effect is slightly reduced when using threshold detectors due to the variance reduction introduced (referred to in section \ref{subsec:theshold}).
 
 At the same time, the width of the 68 \% confidence interval is larger (and spreads more with increasing fluctuations) for the coherent source than for the BinMux-SP source, for both types of detectors. For example, in the NR case, it suffers a fourfold increase for the coherent source, while for the BinMux-SP $(m=5)$ it increases almost 1.5 times.

These results show that the effect of raising the single-photon probability introduced by the time multiplexing stage in the BinMux-SP source, also guarantees a more robust output flux against fluctuations of the pump power intensity. This is mainly due to the fact that less intensity is required at the input to achieve a certain mean number of photons at the output, an effect that increases with increasing number of correcting stages. Taking into account that perfect Poissonian coherent sources (that would require perfectly intensity stable laser sources) are not experimentally abundant, this feature of the BinMux-SP source is also of great importance. 
 
\section{\label{sec:final} Final remarks and discussion}

In this work we propose the use of a specific single-photon source (BinMux-SP) as input for transmission/absorption measurements and we study its performance in obtaining a quantum enhancement in such tasks. 

Perfect number Fock states can achieve the ultimate quantum limit in this type of measurements, but represent a technological challenge, even more so for large numbers. The proposed strategy is to approach an $\nu$ number perfect Fock state using single-photon states generated from the BinMux-SP source as input and performing $\nu$ repetitions. 

We propose a direct measurement scheme, comparing the number of photons detected with and without sample. Due to the single-photon nature of the sources, we have built estimators for both number-resolving and threshold detectors. Working with threshold detectors and achieving sub-shot-noise performance is a desirable goal due to their widespread usage in quantum optics and metrology laboratories. We compare the performance of the BinMux-SP source with that of a perfect single-photon Fock state (UQL) and that of a weak coherent pulse source with Poissonian statistics.

In terms of accuracy, the number-resolving detectors are unbiased, as opposed to the threshold detectors. This states that while the precision may be arbitrarily reduced by the the number of repetitions in the number-resolving case, a fundamental limit exists in the threshold case.  Nonetheless, the bias is non-trivially reduced to zero for $t = 1$, indicating that an acceptable performance can still be obtained for low absorption samples. Explicitly, for a transmission $t = 0.98$, an enhancement factor of 2.2 over the SNL can be achieved with $\nu = 200$ repetitions and $\langle n_b \rangle = 1$ using the BinMux-SP source with threshold detectors. This measurement implies a relative MSE of 5\%.

For other values of transmission, such as $t = 0.8$, in order to reduce the effect of the bias it is convenient to work at even lower mean photon numbers, such as $\langle n_b \rangle = 0.5$. With such figures, an enhancement factor of 1.4 with a relative MSE of 3.1\% can be achieved with $\nu = 2000$ repetitions. 

In the number-resolving case, for a transmissions of $t = 0.98$ and $t = 0.8$, maximum enhancement factors of 1.6 and 1.45 can be achieved respectively, using $\langle n_b \rangle = 0.6$. The relative MSE can be arbitrarily reduced, due to the zero bias of the estimators.  

Taking into account that perfect Poissonian sources are not easy to build, in section \ref{sec:fluctuations} we showed that the performance of a temporally multiplexed single-photon source is less influenced by super-Poissonian fluctuations than that of the coherent source. This is an interesting result for real-life imperfect experimental implementations. 

The results presented in this work encourage the use of single-photon sources as suitable input beams for transmission estimation, while achieving a large quantum enhancement. Particularly, this study of the performance of the BinMux-SP source taking into account a several experimental imperfections shows that single-photon sources built from SPDC and time multiplexing strategies represent a valid, cost-effective and room-temperature alternative to other single-photon sources.

\bibliography{transm_tmux}

\providecommand{\noopsort}[1]{}\providecommand{\singleletter}[1]{#1}%
\begin{thebibliography}{69}%
\makeatletter
\providecommand \@ifxundefined [1]{%
 \@ifx{#1\undefined}
}%
\providecommand \@ifnum [1]{%
 \ifnum #1\expandafter \@firstoftwo
 \else \expandafter \@secondoftwo
 \fi
}%
\providecommand \@ifx [1]{%
 \ifx #1\expandafter \@firstoftwo
 \else \expandafter \@secondoftwo
 \fi
}%
\providecommand \natexlab [1]{#1}%
\providecommand \enquote  [1]{``#1''}%
\providecommand \bibnamefont  [1]{#1}%
\providecommand \bibfnamefont [1]{#1}%
\providecommand \citenamefont [1]{#1}%
\providecommand \href@noop [0]{\@secondoftwo}%
\providecommand \href [0]{\begingroup \@sanitize@url \@href}%
\providecommand \@href[1]{\@@startlink{#1}\@@href}%
\providecommand \@@href[1]{\endgroup#1\@@endlink}%
\providecommand \@sanitize@url [0]{\catcode `\\12\catcode `\$12\catcode
  `\&12\catcode `\#12\catcode `\^12\catcode `\_12\catcode `\%12\relax}%
\providecommand \@@startlink[1]{}%
\providecommand \@@endlink[0]{}%
\providecommand \url  [0]{\begingroup\@sanitize@url \@url }%
\providecommand \@url [1]{\endgroup\@href {#1}{\urlprefix }}%
\providecommand \urlprefix  [0]{URL }%
\providecommand \Eprint [0]{\href }%
\providecommand \doibase [0]{https://doi.org/}%
\providecommand \selectlanguage [0]{\@gobble}%
\providecommand \bibinfo  [0]{\@secondoftwo}%
\providecommand \bibfield  [0]{\@secondoftwo}%
\providecommand \translation [1]{[#1]}%
\providecommand \BibitemOpen [0]{}%
\providecommand \bibitemStop [0]{}%
\providecommand \bibitemNoStop [0]{.\EOS\space}%
\providecommand \EOS [0]{\spacefactor3000\relax}%
\providecommand \BibitemShut  [1]{\csname bibitem#1\endcsname}%
\let\auto@bib@innerbib\@empty
\bibitem [{\citenamefont {Schmiegelow}\ and\ \citenamefont
  {Larotonda}(2014)}]{schmiegelow2014multiplexing}%
  \BibitemOpen
  \bibfield  {author} {\bibinfo {author} {\bibfnamefont {C.~T.}\ \bibnamefont
  {Schmiegelow}}\ and\ \bibinfo {author} {\bibfnamefont {M.~A.}\ \bibnamefont
  {Larotonda}},\ }\bibfield  {title} {\bibinfo {title} {Multiplexing photons
  with a binary division strategy},\ }\href@noop {} {\bibfield  {journal}
  {\bibinfo  {journal} {Applied Physics B}\ }\textbf {\bibinfo {volume}
  {116}},\ \bibinfo {pages} {447} (\bibinfo {year} {2014})}\BibitemShut
  {NoStop}%
\bibitem [{\citenamefont {Giovannetti}\ \emph {et~al.}(2011)\citenamefont
  {Giovannetti}, \citenamefont {Lloyd},\ and\ \citenamefont
  {Maccone}}]{giovanetti2011}%
  \BibitemOpen
  \bibfield  {author} {\bibinfo {author} {\bibfnamefont {V.}~\bibnamefont
  {Giovannetti}}, \bibinfo {author} {\bibfnamefont {S.}~\bibnamefont {Lloyd}},\
  and\ \bibinfo {author} {\bibfnamefont {L.}~\bibnamefont {Maccone}},\
  }\bibfield  {title} {\bibinfo {title} {Advances in quantum metrology},\
  }\href@noop {} {\bibfield  {journal} {\bibinfo  {journal} {Nature Photonics}\
  }\textbf {\bibinfo {volume} {5}},\ \bibinfo {pages} {222–229} (\bibinfo
  {year} {2011})}\BibitemShut {NoStop}%
\bibitem [{\citenamefont {Taylor}\ and\ \citenamefont
  {Bowen}(2016)}]{taylor2016quantum}%
  \BibitemOpen
  \bibfield  {author} {\bibinfo {author} {\bibfnamefont {M.~A.}\ \bibnamefont
  {Taylor}}\ and\ \bibinfo {author} {\bibfnamefont {W.~P.}\ \bibnamefont
  {Bowen}},\ }\bibfield  {title} {\bibinfo {title} {Quantum metrology and its
  application in biology},\ }\href@noop {} {\bibfield  {journal} {\bibinfo
  {journal} {Physics Reports}\ }\textbf {\bibinfo {volume} {615}},\ \bibinfo
  {pages} {1} (\bibinfo {year} {2016})}\BibitemShut {NoStop}%
\bibitem [{\citenamefont {Wasielewski}\ \emph {et~al.}(2020)\citenamefont
  {Wasielewski}, \citenamefont {Forbes}, \citenamefont {Frank}, \citenamefont
  {Kowalski}, \citenamefont {Scholes}, \citenamefont {Yuen-Zhou}, \citenamefont
  {Baldo}, \citenamefont {Freedman}, \citenamefont {Goldsmith}, \citenamefont
  {Goodson} \emph {et~al.}}]{wasielewski2020exploiting}%
  \BibitemOpen
  \bibfield  {author} {\bibinfo {author} {\bibfnamefont {M.~R.}\ \bibnamefont
  {Wasielewski}}, \bibinfo {author} {\bibfnamefont {M.~D.}\ \bibnamefont
  {Forbes}}, \bibinfo {author} {\bibfnamefont {N.~L.}\ \bibnamefont {Frank}},
  \bibinfo {author} {\bibfnamefont {K.}~\bibnamefont {Kowalski}}, \bibinfo
  {author} {\bibfnamefont {G.~D.}\ \bibnamefont {Scholes}}, \bibinfo {author}
  {\bibfnamefont {J.}~\bibnamefont {Yuen-Zhou}}, \bibinfo {author}
  {\bibfnamefont {M.~A.}\ \bibnamefont {Baldo}}, \bibinfo {author}
  {\bibfnamefont {D.~E.}\ \bibnamefont {Freedman}}, \bibinfo {author}
  {\bibfnamefont {R.~H.}\ \bibnamefont {Goldsmith}}, \bibinfo {author}
  {\bibfnamefont {T.}~\bibnamefont {Goodson}}, \emph {et~al.},\ }\bibfield
  {title} {\bibinfo {title} {Exploiting chemistry and molecular systems for
  quantum information science},\ }\href@noop {} {\bibfield  {journal} {\bibinfo
   {journal} {Nature Reviews Chemistry}\ ,\ \bibinfo {pages} {1}} (\bibinfo
  {year} {2020})}\BibitemShut {NoStop}%
\bibitem [{\citenamefont {Maga{\~n}a-Loaiza}\ and\ \citenamefont
  {Boyd}(2019)}]{magana2019quantum}%
  \BibitemOpen
  \bibfield  {author} {\bibinfo {author} {\bibfnamefont {O.~S.}\ \bibnamefont
  {Maga{\~n}a-Loaiza}}\ and\ \bibinfo {author} {\bibfnamefont {R.~W.}\
  \bibnamefont {Boyd}},\ }\bibfield  {title} {\bibinfo {title} {Quantum imaging
  and information},\ }\href@noop {} {\bibfield  {journal} {\bibinfo  {journal}
  {Reports on Progress in Physics}\ }\textbf {\bibinfo {volume} {82}},\
  \bibinfo {pages} {124401} (\bibinfo {year} {2019})}\BibitemShut {NoStop}%
\bibitem [{\citenamefont {Pang}\ and\ \citenamefont
  {Jordan}(2017)}]{pang2017optimal}%
  \BibitemOpen
  \bibfield  {author} {\bibinfo {author} {\bibfnamefont {S.}~\bibnamefont
  {Pang}}\ and\ \bibinfo {author} {\bibfnamefont {A.~N.}\ \bibnamefont
  {Jordan}},\ }\bibfield  {title} {\bibinfo {title} {Optimal adaptive control
  for quantum metrology with time-dependent hamiltonians},\ }\href@noop {}
  {\bibfield  {journal} {\bibinfo  {journal} {Nature communications}\ }\textbf
  {\bibinfo {volume} {8}},\ \bibinfo {pages} {1} (\bibinfo {year}
  {2017})}\BibitemShut {NoStop}%
\bibitem [{\citenamefont {Berchera}\ and\ \citenamefont
  {Degiovanni}(2019)}]{berchera2019quantum}%
  \BibitemOpen
  \bibfield  {author} {\bibinfo {author} {\bibfnamefont {I.~R.}\ \bibnamefont
  {Berchera}}\ and\ \bibinfo {author} {\bibfnamefont {I.~P.}\ \bibnamefont
  {Degiovanni}},\ }\bibfield  {title} {\bibinfo {title} {Quantum imaging with
  sub-poissonian light: challenges and perspectives in optical metrology},\
  }\href@noop {} {\bibfield  {journal} {\bibinfo  {journal} {Metrologia}\
  }\textbf {\bibinfo {volume} {56}},\ \bibinfo {pages} {024001} (\bibinfo
  {year} {2019})}\BibitemShut {NoStop}%
\bibitem [{\citenamefont {Moreau}\ \emph {et~al.}(2019)\citenamefont {Moreau},
  \citenamefont {Toninelli}, \citenamefont {Gregory},\ and\ \citenamefont
  {Padgett}}]{moreau2019imaging}%
  \BibitemOpen
  \bibfield  {author} {\bibinfo {author} {\bibfnamefont {P.-A.}\ \bibnamefont
  {Moreau}}, \bibinfo {author} {\bibfnamefont {E.}~\bibnamefont {Toninelli}},
  \bibinfo {author} {\bibfnamefont {T.}~\bibnamefont {Gregory}},\ and\ \bibinfo
  {author} {\bibfnamefont {M.~J.}\ \bibnamefont {Padgett}},\ }\bibfield
  {title} {\bibinfo {title} {Imaging with quantum states of light},\
  }\href@noop {} {\bibfield  {journal} {\bibinfo  {journal} {Nature Reviews
  Physics}\ }\textbf {\bibinfo {volume} {1}},\ \bibinfo {pages} {367} (\bibinfo
  {year} {2019})}\BibitemShut {NoStop}%
\bibitem [{\citenamefont {Meda}\ \emph {et~al.}(2017)\citenamefont {Meda},
  \citenamefont {Losero}, \citenamefont {Samantaray}, \citenamefont
  {Scafirimuto}, \citenamefont {Pradyumna}, \citenamefont {Avella},
  \citenamefont {Ruo-Berchera},\ and\ \citenamefont
  {Genovese}}]{meda2017photon}%
  \BibitemOpen
  \bibfield  {author} {\bibinfo {author} {\bibfnamefont {A.}~\bibnamefont
  {Meda}}, \bibinfo {author} {\bibfnamefont {E.}~\bibnamefont {Losero}},
  \bibinfo {author} {\bibfnamefont {N.}~\bibnamefont {Samantaray}}, \bibinfo
  {author} {\bibfnamefont {F.}~\bibnamefont {Scafirimuto}}, \bibinfo {author}
  {\bibfnamefont {S.}~\bibnamefont {Pradyumna}}, \bibinfo {author}
  {\bibfnamefont {A.}~\bibnamefont {Avella}}, \bibinfo {author} {\bibfnamefont
  {I.}~\bibnamefont {Ruo-Berchera}},\ and\ \bibinfo {author} {\bibfnamefont
  {M.}~\bibnamefont {Genovese}},\ }\bibfield  {title} {\bibinfo {title}
  {Photon-number correlation for quantum enhanced imaging and sensing},\
  }\href@noop {} {\bibfield  {journal} {\bibinfo  {journal} {Journal of
  Optics}\ }\textbf {\bibinfo {volume} {19}},\ \bibinfo {pages} {094002}
  (\bibinfo {year} {2017})}\BibitemShut {NoStop}%
\bibitem [{\citenamefont {Yoon}\ \emph {et~al.}(2020)\citenamefont {Yoon},
  \citenamefont {Lee}, \citenamefont {Rockstuhl}, \citenamefont {Lee},\ and\
  \citenamefont {Lee}}]{yoon2020experimental}%
  \BibitemOpen
  \bibfield  {author} {\bibinfo {author} {\bibfnamefont {S.-J.}\ \bibnamefont
  {Yoon}}, \bibinfo {author} {\bibfnamefont {J.-S.}\ \bibnamefont {Lee}},
  \bibinfo {author} {\bibfnamefont {C.}~\bibnamefont {Rockstuhl}}, \bibinfo
  {author} {\bibfnamefont {C.}~\bibnamefont {Lee}},\ and\ \bibinfo {author}
  {\bibfnamefont {K.-G.}\ \bibnamefont {Lee}},\ }\bibfield  {title} {\bibinfo
  {title} {Experimental quantum polarimetry using heralded single photons},\
  }\href@noop {} {\bibfield  {journal} {\bibinfo  {journal} {Metrologia}\ }
  (\bibinfo {year} {2020})}\BibitemShut {NoStop}%
\bibitem [{\citenamefont {Mauranyapin}\ \emph {et~al.}(2017)\citenamefont
  {Mauranyapin}, \citenamefont {Madsen}, \citenamefont {Taylor}, \citenamefont
  {Waleed},\ and\ \citenamefont {Bowen}}]{mauranyapin2017evanescent}%
  \BibitemOpen
  \bibfield  {author} {\bibinfo {author} {\bibfnamefont {N.}~\bibnamefont
  {Mauranyapin}}, \bibinfo {author} {\bibfnamefont {L.}~\bibnamefont {Madsen}},
  \bibinfo {author} {\bibfnamefont {M.}~\bibnamefont {Taylor}}, \bibinfo
  {author} {\bibfnamefont {M.}~\bibnamefont {Waleed}},\ and\ \bibinfo {author}
  {\bibfnamefont {W.}~\bibnamefont {Bowen}},\ }\bibfield  {title} {\bibinfo
  {title} {Evanescent single-molecule biosensing with quantum-limited
  precision},\ }\href@noop {} {\bibfield  {journal} {\bibinfo  {journal}
  {Nature Photonics}\ }\textbf {\bibinfo {volume} {11}},\ \bibinfo {pages}
  {477} (\bibinfo {year} {2017})}\BibitemShut {NoStop}%
\bibitem [{\citenamefont {Zonios}\ \emph {et~al.}(2008)\citenamefont {Zonios},
  \citenamefont {Dimou}, \citenamefont {Bassukas}, \citenamefont {Galaris},
  \citenamefont {Tsolakidis},\ and\ \citenamefont
  {Kaxiras}}]{zonios2008melanin}%
  \BibitemOpen
  \bibfield  {author} {\bibinfo {author} {\bibfnamefont {G.}~\bibnamefont
  {Zonios}}, \bibinfo {author} {\bibfnamefont {A.}~\bibnamefont {Dimou}},
  \bibinfo {author} {\bibfnamefont {I.}~\bibnamefont {Bassukas}}, \bibinfo
  {author} {\bibfnamefont {D.}~\bibnamefont {Galaris}}, \bibinfo {author}
  {\bibfnamefont {A.}~\bibnamefont {Tsolakidis}},\ and\ \bibinfo {author}
  {\bibfnamefont {E.}~\bibnamefont {Kaxiras}},\ }\bibfield  {title} {\bibinfo
  {title} {Melanin absorption spectroscopy: new method for noninvasive skin
  investigation and melanoma detection},\ }\href@noop {} {\bibfield  {journal}
  {\bibinfo  {journal} {Journal of biomedical optics}\ }\textbf {\bibinfo
  {volume} {13}},\ \bibinfo {pages} {014017} (\bibinfo {year}
  {2008})}\BibitemShut {NoStop}%
\bibitem [{\citenamefont {Cone}\ \emph {et~al.}(2015)\citenamefont {Cone},
  \citenamefont {Mason}, \citenamefont {Figueroa}, \citenamefont {Hokr},
  \citenamefont {Bixler}, \citenamefont {Castellanos}, \citenamefont {Noojin},
  \citenamefont {Wigle}, \citenamefont {Rockwell}, \citenamefont {Yakovlev}
  \emph {et~al.}}]{cone2015measuring}%
  \BibitemOpen
  \bibfield  {author} {\bibinfo {author} {\bibfnamefont {M.~T.}\ \bibnamefont
  {Cone}}, \bibinfo {author} {\bibfnamefont {J.~D.}\ \bibnamefont {Mason}},
  \bibinfo {author} {\bibfnamefont {E.}~\bibnamefont {Figueroa}}, \bibinfo
  {author} {\bibfnamefont {B.~H.}\ \bibnamefont {Hokr}}, \bibinfo {author}
  {\bibfnamefont {J.~N.}\ \bibnamefont {Bixler}}, \bibinfo {author}
  {\bibfnamefont {C.~C.}\ \bibnamefont {Castellanos}}, \bibinfo {author}
  {\bibfnamefont {G.~D.}\ \bibnamefont {Noojin}}, \bibinfo {author}
  {\bibfnamefont {J.~C.}\ \bibnamefont {Wigle}}, \bibinfo {author}
  {\bibfnamefont {B.~A.}\ \bibnamefont {Rockwell}}, \bibinfo {author}
  {\bibfnamefont {V.~V.}\ \bibnamefont {Yakovlev}}, \emph {et~al.},\ }\bibfield
   {title} {\bibinfo {title} {Measuring the absorption coefficient of
  biological materials using integrating cavity ring-down spectroscopy},\
  }\href@noop {} {\bibfield  {journal} {\bibinfo  {journal} {Optica}\ }\textbf
  {\bibinfo {volume} {2}},\ \bibinfo {pages} {162} (\bibinfo {year}
  {2015})}\BibitemShut {NoStop}%
\bibitem [{\citenamefont {Diedrich}\ and\ \citenamefont
  {Walther}(1987)}]{diedrich1987nonclassical}%
  \BibitemOpen
  \bibfield  {author} {\bibinfo {author} {\bibfnamefont {F.}~\bibnamefont
  {Diedrich}}\ and\ \bibinfo {author} {\bibfnamefont {H.}~\bibnamefont
  {Walther}},\ }\bibfield  {title} {\bibinfo {title} {Nonclassical radiation of
  a single stored ion},\ }\href@noop {} {\bibfield  {journal} {\bibinfo
  {journal} {Physical review letters}\ }\textbf {\bibinfo {volume} {58}},\
  \bibinfo {pages} {203} (\bibinfo {year} {1987})}\BibitemShut {NoStop}%
\bibitem [{\citenamefont {Kimble}\ \emph {et~al.}(1977)\citenamefont {Kimble},
  \citenamefont {Dagenais},\ and\ \citenamefont {Mandel}}]{kimble1977photon}%
  \BibitemOpen
  \bibfield  {author} {\bibinfo {author} {\bibfnamefont {H.~J.}\ \bibnamefont
  {Kimble}}, \bibinfo {author} {\bibfnamefont {M.}~\bibnamefont {Dagenais}},\
  and\ \bibinfo {author} {\bibfnamefont {L.}~\bibnamefont {Mandel}},\
  }\bibfield  {title} {\bibinfo {title} {Photon antibunching in resonance
  fluorescence},\ }\href@noop {} {\bibfield  {journal} {\bibinfo  {journal}
  {Physical Review Letters}\ }\textbf {\bibinfo {volume} {39}},\ \bibinfo
  {pages} {691} (\bibinfo {year} {1977})}\BibitemShut {NoStop}%
\bibitem [{\citenamefont {Mandel}(1979)}]{mandel1979sub}%
  \BibitemOpen
  \bibfield  {author} {\bibinfo {author} {\bibfnamefont {L.}~\bibnamefont
  {Mandel}},\ }\bibfield  {title} {\bibinfo {title} {Sub-poissonian photon
  statistics in resonance fluorescence},\ }\href@noop {} {\bibfield  {journal}
  {\bibinfo  {journal} {Optics letters}\ }\textbf {\bibinfo {volume} {4}},\
  \bibinfo {pages} {205} (\bibinfo {year} {1979})}\BibitemShut {NoStop}%
\bibitem [{\citenamefont {Basch{\'e}}\ \emph {et~al.}(1992)\citenamefont
  {Basch{\'e}}, \citenamefont {Moerner}, \citenamefont {Orrit},\ and\
  \citenamefont {Talon}}]{basche1992photon}%
  \BibitemOpen
  \bibfield  {author} {\bibinfo {author} {\bibfnamefont {T.}~\bibnamefont
  {Basch{\'e}}}, \bibinfo {author} {\bibfnamefont {W.}~\bibnamefont {Moerner}},
  \bibinfo {author} {\bibfnamefont {M.}~\bibnamefont {Orrit}},\ and\ \bibinfo
  {author} {\bibfnamefont {H.}~\bibnamefont {Talon}},\ }\bibfield  {title}
  {\bibinfo {title} {Photon antibunching in the fluorescence of a single dye
  molecule trapped in a solid},\ }\href@noop {} {\bibfield  {journal} {\bibinfo
   {journal} {Physical review letters}\ }\textbf {\bibinfo {volume} {69}},\
  \bibinfo {pages} {1516} (\bibinfo {year} {1992})}\BibitemShut {NoStop}%
\bibitem [{\citenamefont {Aharonovich}\ \emph {et~al.}(2016)\citenamefont
  {Aharonovich}, \citenamefont {Englund},\ and\ \citenamefont
  {Toth}}]{aharonovich2016solid}%
  \BibitemOpen
  \bibfield  {author} {\bibinfo {author} {\bibfnamefont {I.}~\bibnamefont
  {Aharonovich}}, \bibinfo {author} {\bibfnamefont {D.}~\bibnamefont
  {Englund}},\ and\ \bibinfo {author} {\bibfnamefont {M.}~\bibnamefont
  {Toth}},\ }\bibfield  {title} {\bibinfo {title} {Solid-state single-photon
  emitters},\ }\href@noop {} {\bibfield  {journal} {\bibinfo  {journal} {Nature
  Photonics}\ }\textbf {\bibinfo {volume} {10}},\ \bibinfo {pages} {631}
  (\bibinfo {year} {2016})}\BibitemShut {NoStop}%
\bibitem [{\citenamefont {Couteau}\ \emph {et~al.}(2004)\citenamefont
  {Couteau}, \citenamefont {Moehl}, \citenamefont {Tinjod}, \citenamefont
  {G{\'e}rard}, \citenamefont {Kheng}, \citenamefont {Mariette}, \citenamefont
  {Gaj}, \citenamefont {Romestain},\ and\ \citenamefont
  {Poizat}}]{couteau2004correlated}%
  \BibitemOpen
  \bibfield  {author} {\bibinfo {author} {\bibfnamefont {C.}~\bibnamefont
  {Couteau}}, \bibinfo {author} {\bibfnamefont {S.}~\bibnamefont {Moehl}},
  \bibinfo {author} {\bibfnamefont {F.}~\bibnamefont {Tinjod}}, \bibinfo
  {author} {\bibfnamefont {J.}~\bibnamefont {G{\'e}rard}}, \bibinfo {author}
  {\bibfnamefont {K.}~\bibnamefont {Kheng}}, \bibinfo {author} {\bibfnamefont
  {H.}~\bibnamefont {Mariette}}, \bibinfo {author} {\bibfnamefont
  {J.}~\bibnamefont {Gaj}}, \bibinfo {author} {\bibfnamefont {R.}~\bibnamefont
  {Romestain}},\ and\ \bibinfo {author} {\bibfnamefont {J.}~\bibnamefont
  {Poizat}},\ }\bibfield  {title} {\bibinfo {title} {Correlated photon emission
  from a single {II}--{VI} quantum dot},\ }\href@noop {} {\bibfield  {journal}
  {\bibinfo  {journal} {Applied Physics Letters}\ }\textbf {\bibinfo {volume}
  {85}},\ \bibinfo {pages} {6251} (\bibinfo {year} {2004})}\BibitemShut
  {NoStop}%
\bibitem [{\citenamefont {Holmes}\ \emph {et~al.}(2014)\citenamefont {Holmes},
  \citenamefont {Choi}, \citenamefont {Kako}, \citenamefont {Arita},\ and\
  \citenamefont {Arakawa}}]{holmes2014room}%
  \BibitemOpen
  \bibfield  {author} {\bibinfo {author} {\bibfnamefont {M.~J.}\ \bibnamefont
  {Holmes}}, \bibinfo {author} {\bibfnamefont {K.}~\bibnamefont {Choi}},
  \bibinfo {author} {\bibfnamefont {S.}~\bibnamefont {Kako}}, \bibinfo {author}
  {\bibfnamefont {M.}~\bibnamefont {Arita}},\ and\ \bibinfo {author}
  {\bibfnamefont {Y.}~\bibnamefont {Arakawa}},\ }\bibfield  {title} {\bibinfo
  {title} {Room-temperature triggered single photon emission from a
  {III}-nitride site-controlled nanowire quantum dot},\ }\href@noop {}
  {\bibfield  {journal} {\bibinfo  {journal} {Nano letters}\ }\textbf {\bibinfo
  {volume} {14}},\ \bibinfo {pages} {982} (\bibinfo {year} {2014})}\BibitemShut
  {NoStop}%
\bibitem [{\citenamefont {Sebald}\ \emph {et~al.}(2002)\citenamefont {Sebald},
  \citenamefont {Michler}, \citenamefont {Passow}, \citenamefont {Hommel},
  \citenamefont {Bacher},\ and\ \citenamefont {Forchel}}]{sebald2002single}%
  \BibitemOpen
  \bibfield  {author} {\bibinfo {author} {\bibfnamefont {K.}~\bibnamefont
  {Sebald}}, \bibinfo {author} {\bibfnamefont {P.}~\bibnamefont {Michler}},
  \bibinfo {author} {\bibfnamefont {T.}~\bibnamefont {Passow}}, \bibinfo
  {author} {\bibfnamefont {D.}~\bibnamefont {Hommel}}, \bibinfo {author}
  {\bibfnamefont {G.}~\bibnamefont {Bacher}},\ and\ \bibinfo {author}
  {\bibfnamefont {A.}~\bibnamefont {Forchel}},\ }\bibfield  {title} {\bibinfo
  {title} {Single-photon emission of {CdSe} quantum dots at temperatures up to
  200 {K}},\ }\href@noop {} {\bibfield  {journal} {\bibinfo  {journal} {Applied
  Physics Letters}\ }\textbf {\bibinfo {volume} {81}},\ \bibinfo {pages} {2920}
  (\bibinfo {year} {2002})}\BibitemShut {NoStop}%
\bibitem [{\citenamefont {Gammon}\ \emph {et~al.}(1996)\citenamefont {Gammon},
  \citenamefont {Snow}, \citenamefont {Shanabrook}, \citenamefont {Katzer},\
  and\ \citenamefont {Park}}]{gammon1996homogeneous}%
  \BibitemOpen
  \bibfield  {author} {\bibinfo {author} {\bibfnamefont {D.}~\bibnamefont
  {Gammon}}, \bibinfo {author} {\bibfnamefont {E.}~\bibnamefont {Snow}},
  \bibinfo {author} {\bibfnamefont {B.}~\bibnamefont {Shanabrook}}, \bibinfo
  {author} {\bibfnamefont {D.}~\bibnamefont {Katzer}},\ and\ \bibinfo {author}
  {\bibfnamefont {D.}~\bibnamefont {Park}},\ }\bibfield  {title} {\bibinfo
  {title} {Homogeneous linewidths in the optical spectrum of a single gallium
  arsenide quantum dot},\ }\href@noop {} {\bibfield  {journal} {\bibinfo
  {journal} {Science}\ }\textbf {\bibinfo {volume} {273}},\ \bibinfo {pages}
  {87} (\bibinfo {year} {1996})}\BibitemShut {NoStop}%
\bibitem [{\citenamefont {Michler}\ \emph {et~al.}(2000)\citenamefont
  {Michler}, \citenamefont {Kiraz}, \citenamefont {Becher}, \citenamefont
  {Schoenfeld}, \citenamefont {Petroff}, \citenamefont {Zhang}, \citenamefont
  {Hu},\ and\ \citenamefont {Imamoglu}}]{michler2000quantum}%
  \BibitemOpen
  \bibfield  {author} {\bibinfo {author} {\bibfnamefont {P.}~\bibnamefont
  {Michler}}, \bibinfo {author} {\bibfnamefont {A.}~\bibnamefont {Kiraz}},
  \bibinfo {author} {\bibfnamefont {C.}~\bibnamefont {Becher}}, \bibinfo
  {author} {\bibfnamefont {W.}~\bibnamefont {Schoenfeld}}, \bibinfo {author}
  {\bibfnamefont {P.}~\bibnamefont {Petroff}}, \bibinfo {author} {\bibfnamefont
  {L.}~\bibnamefont {Zhang}}, \bibinfo {author} {\bibfnamefont
  {E.}~\bibnamefont {Hu}},\ and\ \bibinfo {author} {\bibfnamefont
  {A.}~\bibnamefont {Imamoglu}},\ }\bibfield  {title} {\bibinfo {title} {A
  quantum dot single-photon turnstile device},\ }\href@noop {} {\bibfield
  {journal} {\bibinfo  {journal} {Science}\ }\textbf {\bibinfo {volume}
  {290}},\ \bibinfo {pages} {2282} (\bibinfo {year} {2000})}\BibitemShut
  {NoStop}%
\bibitem [{\citenamefont {Ding}\ \emph {et~al.}(2016)\citenamefont {Ding},
  \citenamefont {He}, \citenamefont {Duan}, \citenamefont {Gregersen},
  \citenamefont {Chen}, \citenamefont {Unsleber}, \citenamefont {Maier},
  \citenamefont {Schneider}, \citenamefont {Kamp}, \citenamefont {H\"ofling},
  \citenamefont {Lu},\ and\ \citenamefont {Pan}}]{QDpan}%
  \BibitemOpen
  \bibfield  {author} {\bibinfo {author} {\bibfnamefont {X.}~\bibnamefont
  {Ding}}, \bibinfo {author} {\bibfnamefont {Y.}~\bibnamefont {He}}, \bibinfo
  {author} {\bibfnamefont {Z.-C.}\ \bibnamefont {Duan}}, \bibinfo {author}
  {\bibfnamefont {N.}~\bibnamefont {Gregersen}}, \bibinfo {author}
  {\bibfnamefont {M.-C.}\ \bibnamefont {Chen}}, \bibinfo {author}
  {\bibfnamefont {S.}~\bibnamefont {Unsleber}}, \bibinfo {author}
  {\bibfnamefont {S.}~\bibnamefont {Maier}}, \bibinfo {author} {\bibfnamefont
  {C.}~\bibnamefont {Schneider}}, \bibinfo {author} {\bibfnamefont
  {M.}~\bibnamefont {Kamp}}, \bibinfo {author} {\bibfnamefont {S.}~\bibnamefont
  {H\"ofling}}, \bibinfo {author} {\bibfnamefont {C.-Y.}\ \bibnamefont {Lu}},\
  and\ \bibinfo {author} {\bibfnamefont {J.-W.}\ \bibnamefont {Pan}},\
  }\bibfield  {title} {\bibinfo {title} {On-demand single photons with high
  extraction efficiency and near-unity indistinguishability from a resonantly
  driven quantum dot in a micropillar},\ }\href
  {https://doi.org/10.1103/PhysRevLett.116.020401} {\bibfield  {journal}
  {\bibinfo  {journal} {Phys. Rev. Lett.}\ }\textbf {\bibinfo {volume} {116}},\
  \bibinfo {pages} {020401} (\bibinfo {year} {2016})}\BibitemShut {NoStop}%
\bibitem [{\citenamefont {Somaschi}\ \emph {et~al.}(2016)\citenamefont
  {Somaschi}, \citenamefont {Giesz}, \citenamefont {De~Santis}, \citenamefont
  {Loredo}, \citenamefont {Almeida}, \citenamefont {Hornecker}, \citenamefont
  {Portalupi}, \citenamefont {Grange}, \citenamefont {Ant{\'o}n}, \citenamefont
  {Demory} \emph {et~al.}}]{somaschi2016near}%
  \BibitemOpen
  \bibfield  {author} {\bibinfo {author} {\bibfnamefont {N.}~\bibnamefont
  {Somaschi}}, \bibinfo {author} {\bibfnamefont {V.}~\bibnamefont {Giesz}},
  \bibinfo {author} {\bibfnamefont {L.}~\bibnamefont {De~Santis}}, \bibinfo
  {author} {\bibfnamefont {J.}~\bibnamefont {Loredo}}, \bibinfo {author}
  {\bibfnamefont {M.~P.}\ \bibnamefont {Almeida}}, \bibinfo {author}
  {\bibfnamefont {G.}~\bibnamefont {Hornecker}}, \bibinfo {author}
  {\bibfnamefont {S.~L.}\ \bibnamefont {Portalupi}}, \bibinfo {author}
  {\bibfnamefont {T.}~\bibnamefont {Grange}}, \bibinfo {author} {\bibfnamefont
  {C.}~\bibnamefont {Ant{\'o}n}}, \bibinfo {author} {\bibfnamefont
  {J.}~\bibnamefont {Demory}}, \emph {et~al.},\ }\bibfield  {title} {\bibinfo
  {title} {Near-optimal single-photon sources in the solid state},\ }\href@noop
  {} {\bibfield  {journal} {\bibinfo  {journal} {Nature Photonics}\ }\textbf
  {\bibinfo {volume} {10}},\ \bibinfo {pages} {340} (\bibinfo {year}
  {2016})}\BibitemShut {NoStop}%
\bibitem [{\citenamefont {Ollivier}\ \emph {et~al.}(2020)\citenamefont
  {Ollivier}, \citenamefont {Thomas}, \citenamefont {Wein}, \citenamefont
  {Wenniger}, \citenamefont {Coste}, \citenamefont {Loredo}, \citenamefont
  {Somaschi}, \citenamefont {Harouri}, \citenamefont {Lemaitre}, \citenamefont
  {Sagnes} \emph {et~al.}}]{ollivier2020hong}%
  \BibitemOpen
  \bibfield  {author} {\bibinfo {author} {\bibfnamefont {H.}~\bibnamefont
  {Ollivier}}, \bibinfo {author} {\bibfnamefont {S.}~\bibnamefont {Thomas}},
  \bibinfo {author} {\bibfnamefont {S.}~\bibnamefont {Wein}}, \bibinfo {author}
  {\bibfnamefont {I.}~\bibnamefont {Wenniger}}, \bibinfo {author}
  {\bibfnamefont {N.}~\bibnamefont {Coste}}, \bibinfo {author} {\bibfnamefont
  {J.}~\bibnamefont {Loredo}}, \bibinfo {author} {\bibfnamefont
  {N.}~\bibnamefont {Somaschi}}, \bibinfo {author} {\bibfnamefont
  {A.}~\bibnamefont {Harouri}}, \bibinfo {author} {\bibfnamefont
  {A.}~\bibnamefont {Lemaitre}}, \bibinfo {author} {\bibfnamefont
  {I.}~\bibnamefont {Sagnes}}, \emph {et~al.},\ }\bibfield  {title} {\bibinfo
  {title} {Hong-ou-mandel interference with imperfect single photon sources},\
  }\href@noop {} {\bibfield  {journal} {\bibinfo  {journal} {arXiv preprint
  arXiv:2005.01743}\ } (\bibinfo {year} {2020})}\BibitemShut {NoStop}%
\bibitem [{\citenamefont {Bock}\ \emph {et~al.}(2016)\citenamefont {Bock},
  \citenamefont {Lenhard}, \citenamefont {Chunnilall},\ and\ \citenamefont
  {Becher}}]{bock2016highly}%
  \BibitemOpen
  \bibfield  {author} {\bibinfo {author} {\bibfnamefont {M.}~\bibnamefont
  {Bock}}, \bibinfo {author} {\bibfnamefont {A.}~\bibnamefont {Lenhard}},
  \bibinfo {author} {\bibfnamefont {C.}~\bibnamefont {Chunnilall}},\ and\
  \bibinfo {author} {\bibfnamefont {C.}~\bibnamefont {Becher}},\ }\bibfield
  {title} {\bibinfo {title} {Highly efficient heralded single-photon source for
  telecom wavelengths based on a {PPLN} waveguide},\ }\href@noop {} {\bibfield
  {journal} {\bibinfo  {journal} {Optics express}\ }\textbf {\bibinfo {volume}
  {24}},\ \bibinfo {pages} {23992} (\bibinfo {year} {2016})}\BibitemShut
  {NoStop}%
\bibitem [{\citenamefont {Montaut}\ \emph {et~al.}(2017)\citenamefont
  {Montaut}, \citenamefont {Sansoni}, \citenamefont {Meyer-Scott},
  \citenamefont {Ricken}, \citenamefont {Quiring}, \citenamefont {Herrmann},\
  and\ \citenamefont {Silberhorn}}]{montaut2017high}%
  \BibitemOpen
  \bibfield  {author} {\bibinfo {author} {\bibfnamefont {N.}~\bibnamefont
  {Montaut}}, \bibinfo {author} {\bibfnamefont {L.}~\bibnamefont {Sansoni}},
  \bibinfo {author} {\bibfnamefont {E.}~\bibnamefont {Meyer-Scott}}, \bibinfo
  {author} {\bibfnamefont {R.}~\bibnamefont {Ricken}}, \bibinfo {author}
  {\bibfnamefont {V.}~\bibnamefont {Quiring}}, \bibinfo {author} {\bibfnamefont
  {H.}~\bibnamefont {Herrmann}},\ and\ \bibinfo {author} {\bibfnamefont
  {C.}~\bibnamefont {Silberhorn}},\ }\bibfield  {title} {\bibinfo {title}
  {High-efficiency plug-and-play source of heralded single photons},\
  }\href@noop {} {\bibfield  {journal} {\bibinfo  {journal} {Physical Review
  Applied}\ }\textbf {\bibinfo {volume} {8}},\ \bibinfo {pages} {024021}
  (\bibinfo {year} {2017})}\BibitemShut {NoStop}%
\bibitem [{\citenamefont {Paesani}\ \emph {et~al.}(2020)\citenamefont
  {Paesani}, \citenamefont {Borghi}, \citenamefont {Signorini}, \citenamefont
  {Ma{\"\i}nos}, \citenamefont {Pavesi},\ and\ \citenamefont
  {Laing}}]{paesani2020near}%
  \BibitemOpen
  \bibfield  {author} {\bibinfo {author} {\bibfnamefont {S.}~\bibnamefont
  {Paesani}}, \bibinfo {author} {\bibfnamefont {M.}~\bibnamefont {Borghi}},
  \bibinfo {author} {\bibfnamefont {S.}~\bibnamefont {Signorini}}, \bibinfo
  {author} {\bibfnamefont {A.}~\bibnamefont {Ma{\"\i}nos}}, \bibinfo {author}
  {\bibfnamefont {L.}~\bibnamefont {Pavesi}},\ and\ \bibinfo {author}
  {\bibfnamefont {A.}~\bibnamefont {Laing}},\ }\bibfield  {title} {\bibinfo
  {title} {Near-ideal spontaneous photon sources in silicon quantum
  photonics},\ }\href@noop {} {\bibfield  {journal} {\bibinfo  {journal}
  {Nature communications}\ }\textbf {\bibinfo {volume} {11}},\ \bibinfo {pages}
  {1} (\bibinfo {year} {2020})}\BibitemShut {NoStop}%
\bibitem [{\citenamefont {Migdall}\ \emph {et~al.}(2002)\citenamefont
  {Migdall}, \citenamefont {Branning},\ and\ \citenamefont
  {Castelletto}}]{migdall2002tailoring}%
  \BibitemOpen
  \bibfield  {author} {\bibinfo {author} {\bibfnamefont {A.~L.}\ \bibnamefont
  {Migdall}}, \bibinfo {author} {\bibfnamefont {D.}~\bibnamefont {Branning}},\
  and\ \bibinfo {author} {\bibfnamefont {S.}~\bibnamefont {Castelletto}},\
  }\bibfield  {title} {\bibinfo {title} {Tailoring single-photon and
  multiphoton probabilities of a single-photon on-demand source},\ }\href@noop
  {} {\bibfield  {journal} {\bibinfo  {journal} {Physical Review A}\ }\textbf
  {\bibinfo {volume} {66}},\ \bibinfo {pages} {053805} (\bibinfo {year}
  {2002})}\BibitemShut {NoStop}%
\bibitem [{\citenamefont {Shapiro}\ and\ \citenamefont
  {Wong}(2007)}]{shapiro2007demand}%
  \BibitemOpen
  \bibfield  {author} {\bibinfo {author} {\bibfnamefont {J.~H.}\ \bibnamefont
  {Shapiro}}\ and\ \bibinfo {author} {\bibfnamefont {F.~N.}\ \bibnamefont
  {Wong}},\ }\bibfield  {title} {\bibinfo {title} {On-demand single-photon
  generation using a modular array of parametric downconverters with
  electro-optic polarization controls},\ }\href@noop {} {\bibfield  {journal}
  {\bibinfo  {journal} {Optics letters}\ }\textbf {\bibinfo {volume} {32}},\
  \bibinfo {pages} {2698} (\bibinfo {year} {2007})}\BibitemShut {NoStop}%
\bibitem [{\citenamefont {Jennewein}\ \emph {et~al.}(2011)\citenamefont
  {Jennewein}, \citenamefont {Barbieri},\ and\ \citenamefont
  {White}}]{jennewein2011single}%
  \BibitemOpen
  \bibfield  {author} {\bibinfo {author} {\bibfnamefont {T.}~\bibnamefont
  {Jennewein}}, \bibinfo {author} {\bibfnamefont {M.}~\bibnamefont
  {Barbieri}},\ and\ \bibinfo {author} {\bibfnamefont {A.~G.}\ \bibnamefont
  {White}},\ }\bibfield  {title} {\bibinfo {title} {Single-photon device
  requirements for operating linear optics quantum computing outside the
  post-selection basis},\ }\href@noop {} {\bibfield  {journal} {\bibinfo
  {journal} {Journal of Modern Optics}\ }\textbf {\bibinfo {volume} {58}},\
  \bibinfo {pages} {276} (\bibinfo {year} {2011})}\BibitemShut {NoStop}%
\bibitem [{\citenamefont {Ma}\ \emph {et~al.}(2011)\citenamefont {Ma},
  \citenamefont {Zotter}, \citenamefont {Kofler}, \citenamefont {Jennewein},\
  and\ \citenamefont {Zeilinger}}]{ma2011experimental}%
  \BibitemOpen
  \bibfield  {author} {\bibinfo {author} {\bibfnamefont {X.-s.}\ \bibnamefont
  {Ma}}, \bibinfo {author} {\bibfnamefont {S.}~\bibnamefont {Zotter}}, \bibinfo
  {author} {\bibfnamefont {J.}~\bibnamefont {Kofler}}, \bibinfo {author}
  {\bibfnamefont {T.}~\bibnamefont {Jennewein}},\ and\ \bibinfo {author}
  {\bibfnamefont {A.}~\bibnamefont {Zeilinger}},\ }\bibfield  {title} {\bibinfo
  {title} {Experimental generation of single photons via active multiplexing},\
  }\href@noop {} {\bibfield  {journal} {\bibinfo  {journal} {Physical Review
  A}\ }\textbf {\bibinfo {volume} {83}},\ \bibinfo {pages} {043814} (\bibinfo
  {year} {2011})}\BibitemShut {NoStop}%
\bibitem [{\citenamefont {Christ}\ and\ \citenamefont
  {Silberhorn}(2012)}]{christ2012limits}%
  \BibitemOpen
  \bibfield  {author} {\bibinfo {author} {\bibfnamefont {A.}~\bibnamefont
  {Christ}}\ and\ \bibinfo {author} {\bibfnamefont {C.}~\bibnamefont
  {Silberhorn}},\ }\bibfield  {title} {\bibinfo {title} {Limits on the
  deterministic creation of pure single-photon states using parametric
  down-conversion},\ }\href@noop {} {\bibfield  {journal} {\bibinfo  {journal}
  {Physical Review A}\ }\textbf {\bibinfo {volume} {85}},\ \bibinfo {pages}
  {023829} (\bibinfo {year} {2012})}\BibitemShut {NoStop}%
\bibitem [{\citenamefont {Collins}\ \emph {et~al.}(2013)\citenamefont
  {Collins}, \citenamefont {Xiong}, \citenamefont {Rey}, \citenamefont {Vo},
  \citenamefont {He}, \citenamefont {Shahnia}, \citenamefont {Reardon},
  \citenamefont {Krauss}, \citenamefont {Steel}, \citenamefont {Clark} \emph
  {et~al.}}]{collins2013integrated}%
  \BibitemOpen
  \bibfield  {author} {\bibinfo {author} {\bibfnamefont {M.~J.}\ \bibnamefont
  {Collins}}, \bibinfo {author} {\bibfnamefont {C.}~\bibnamefont {Xiong}},
  \bibinfo {author} {\bibfnamefont {I.~H.}\ \bibnamefont {Rey}}, \bibinfo
  {author} {\bibfnamefont {T.~D.}\ \bibnamefont {Vo}}, \bibinfo {author}
  {\bibfnamefont {J.}~\bibnamefont {He}}, \bibinfo {author} {\bibfnamefont
  {S.}~\bibnamefont {Shahnia}}, \bibinfo {author} {\bibfnamefont
  {C.}~\bibnamefont {Reardon}}, \bibinfo {author} {\bibfnamefont {T.~F.}\
  \bibnamefont {Krauss}}, \bibinfo {author} {\bibfnamefont {M.}~\bibnamefont
  {Steel}}, \bibinfo {author} {\bibfnamefont {A.~S.}\ \bibnamefont {Clark}},
  \emph {et~al.},\ }\bibfield  {title} {\bibinfo {title} {Integrated spatial
  multiplexing of heralded single-photon sources},\ }\href@noop {} {\bibfield
  {journal} {\bibinfo  {journal} {Nature communications}\ }\textbf {\bibinfo
  {volume} {4}},\ \bibinfo {pages} {1} (\bibinfo {year} {2013})}\BibitemShut
  {NoStop}%
\bibitem [{\citenamefont {Mazzarella}\ \emph {et~al.}(2013)\citenamefont
  {Mazzarella}, \citenamefont {Ticozzi}, \citenamefont {Sergienko},
  \citenamefont {Vallone},\ and\ \citenamefont
  {Villoresi}}]{mazzarella2013asymmetric}%
  \BibitemOpen
  \bibfield  {author} {\bibinfo {author} {\bibfnamefont {L.}~\bibnamefont
  {Mazzarella}}, \bibinfo {author} {\bibfnamefont {F.}~\bibnamefont {Ticozzi}},
  \bibinfo {author} {\bibfnamefont {A.~V.}\ \bibnamefont {Sergienko}}, \bibinfo
  {author} {\bibfnamefont {G.}~\bibnamefont {Vallone}},\ and\ \bibinfo {author}
  {\bibfnamefont {P.}~\bibnamefont {Villoresi}},\ }\bibfield  {title} {\bibinfo
  {title} {Asymmetric architecture for heralded single-photon sources},\
  }\href@noop {} {\bibfield  {journal} {\bibinfo  {journal} {Physical Review
  A}\ }\textbf {\bibinfo {volume} {88}},\ \bibinfo {pages} {023848} (\bibinfo
  {year} {2013})}\BibitemShut {NoStop}%
\bibitem [{\citenamefont {Jeffrey}\ \emph {et~al.}(2004)\citenamefont
  {Jeffrey}, \citenamefont {Peters},\ and\ \citenamefont
  {Kwiat}}]{jeffrey2004towards}%
  \BibitemOpen
  \bibfield  {author} {\bibinfo {author} {\bibfnamefont {E.}~\bibnamefont
  {Jeffrey}}, \bibinfo {author} {\bibfnamefont {N.~A.}\ \bibnamefont
  {Peters}},\ and\ \bibinfo {author} {\bibfnamefont {P.~G.}\ \bibnamefont
  {Kwiat}},\ }\bibfield  {title} {\bibinfo {title} {Towards a periodic
  deterministic source of arbitrary single-photon states},\ }\href@noop {}
  {\bibfield  {journal} {\bibinfo  {journal} {New Journal of Physics}\ }\textbf
  {\bibinfo {volume} {6}},\ \bibinfo {pages} {100} (\bibinfo {year}
  {2004})}\BibitemShut {NoStop}%
\bibitem [{\citenamefont {Mower}\ and\ \citenamefont
  {Englund}(2011)}]{mower2011efficient}%
  \BibitemOpen
  \bibfield  {author} {\bibinfo {author} {\bibfnamefont {J.}~\bibnamefont
  {Mower}}\ and\ \bibinfo {author} {\bibfnamefont {D.}~\bibnamefont
  {Englund}},\ }\bibfield  {title} {\bibinfo {title} {Efficient generation of
  single and entangled photons on a silicon photonic integrated chip},\
  }\href@noop {} {\bibfield  {journal} {\bibinfo  {journal} {Physical Review
  A}\ }\textbf {\bibinfo {volume} {84}},\ \bibinfo {pages} {052326} (\bibinfo
  {year} {2011})}\BibitemShut {NoStop}%
\bibitem [{\citenamefont {Glebov}\ \emph {et~al.}(2013)\citenamefont {Glebov},
  \citenamefont {Fan},\ and\ \citenamefont
  {Migdall}}]{glebov2013deterministic}%
  \BibitemOpen
  \bibfield  {author} {\bibinfo {author} {\bibfnamefont {B.~L.}\ \bibnamefont
  {Glebov}}, \bibinfo {author} {\bibfnamefont {J.}~\bibnamefont {Fan}},\ and\
  \bibinfo {author} {\bibfnamefont {A.}~\bibnamefont {Migdall}},\ }\bibfield
  {title} {\bibinfo {title} {Deterministic generation of single photons via
  multiplexing repetitive parametric downconversions},\ }\href@noop {}
  {\bibfield  {journal} {\bibinfo  {journal} {Applied Physics Letters}\
  }\textbf {\bibinfo {volume} {103}},\ \bibinfo {pages} {031115} (\bibinfo
  {year} {2013})}\BibitemShut {NoStop}%
\bibitem [{\citenamefont {Kaneda}\ \emph {et~al.}(2015)\citenamefont {Kaneda},
  \citenamefont {Christensen}, \citenamefont {Wong}, \citenamefont {Park},
  \citenamefont {McCusker},\ and\ \citenamefont {Kwiat}}]{kaneda2015time}%
  \BibitemOpen
  \bibfield  {author} {\bibinfo {author} {\bibfnamefont {F.}~\bibnamefont
  {Kaneda}}, \bibinfo {author} {\bibfnamefont {B.~G.}\ \bibnamefont
  {Christensen}}, \bibinfo {author} {\bibfnamefont {J.~J.}\ \bibnamefont
  {Wong}}, \bibinfo {author} {\bibfnamefont {H.~S.}\ \bibnamefont {Park}},
  \bibinfo {author} {\bibfnamefont {K.~T.}\ \bibnamefont {McCusker}},\ and\
  \bibinfo {author} {\bibfnamefont {P.~G.}\ \bibnamefont {Kwiat}},\ }\bibfield
  {title} {\bibinfo {title} {Time-multiplexed heralded single-photon source},\
  }\href@noop {} {\bibfield  {journal} {\bibinfo  {journal} {Optica}\ }\textbf
  {\bibinfo {volume} {2}},\ \bibinfo {pages} {1010} (\bibinfo {year}
  {2015})}\BibitemShut {NoStop}%
\bibitem [{\citenamefont {Mendoza}\ \emph {et~al.}(2016)\citenamefont
  {Mendoza}, \citenamefont {Santagati}, \citenamefont {Munns}, \citenamefont
  {Hemsley}, \citenamefont {Piekarek}, \citenamefont {Mart\'{i}n-L\'{o}pez},
  \citenamefont {Marshall}, \citenamefont {Bonneau}, \citenamefont {Thompson},\
  and\ \citenamefont {O'Brien}}]{mendoza2015active}%
  \BibitemOpen
  \bibfield  {author} {\bibinfo {author} {\bibfnamefont {G.~J.}\ \bibnamefont
  {Mendoza}}, \bibinfo {author} {\bibfnamefont {R.}~\bibnamefont {Santagati}},
  \bibinfo {author} {\bibfnamefont {J.}~\bibnamefont {Munns}}, \bibinfo
  {author} {\bibfnamefont {E.}~\bibnamefont {Hemsley}}, \bibinfo {author}
  {\bibfnamefont {M.}~\bibnamefont {Piekarek}}, \bibinfo {author}
  {\bibfnamefont {E.}~\bibnamefont {Mart\'{i}n-L\'{o}pez}}, \bibinfo {author}
  {\bibfnamefont {G.~D.}\ \bibnamefont {Marshall}}, \bibinfo {author}
  {\bibfnamefont {D.}~\bibnamefont {Bonneau}}, \bibinfo {author} {\bibfnamefont
  {M.~G.}\ \bibnamefont {Thompson}},\ and\ \bibinfo {author} {\bibfnamefont
  {J.~L.}\ \bibnamefont {O'Brien}},\ }\bibfield  {title} {\bibinfo {title}
  {Active temporal and spatial multiplexing of photons},\ }\href
  {https://doi.org/10.1364/OPTICA.3.000127} {\bibfield  {journal} {\bibinfo
  {journal} {Optica}\ }\textbf {\bibinfo {volume} {3}},\ \bibinfo {pages} {127}
  (\bibinfo {year} {2016})}\BibitemShut {NoStop}%
\bibitem [{\citenamefont {Rohde}\ \emph {et~al.}(2015)\citenamefont {Rohde},
  \citenamefont {Helt}, \citenamefont {Steel},\ and\ \citenamefont
  {Gilchrist}}]{rohde2015multiplexed}%
  \BibitemOpen
  \bibfield  {author} {\bibinfo {author} {\bibfnamefont {P.~P.}\ \bibnamefont
  {Rohde}}, \bibinfo {author} {\bibfnamefont {L.}~\bibnamefont {Helt}},
  \bibinfo {author} {\bibfnamefont {M.}~\bibnamefont {Steel}},\ and\ \bibinfo
  {author} {\bibfnamefont {A.}~\bibnamefont {Gilchrist}},\ }\bibfield  {title}
  {\bibinfo {title} {Multiplexed single-photon-state preparation using a
  fiber-loop architecture},\ }\href@noop {} {\bibfield  {journal} {\bibinfo
  {journal} {Physical Review A}\ }\textbf {\bibinfo {volume} {92}},\ \bibinfo
  {pages} {053829} (\bibinfo {year} {2015})}\BibitemShut {NoStop}%
\bibitem [{\citenamefont {Zhang}\ \emph {et~al.}(2017)\citenamefont {Zhang},
  \citenamefont {Lee}, \citenamefont {Bell}, \citenamefont {Leong},
  \citenamefont {Rudolph}, \citenamefont {Eggleton},\ and\ \citenamefont
  {Xiong}}]{zhang2017indistinguishable}%
  \BibitemOpen
  \bibfield  {author} {\bibinfo {author} {\bibfnamefont {X.}~\bibnamefont
  {Zhang}}, \bibinfo {author} {\bibfnamefont {Y.}~\bibnamefont {Lee}}, \bibinfo
  {author} {\bibfnamefont {B.}~\bibnamefont {Bell}}, \bibinfo {author}
  {\bibfnamefont {P.}~\bibnamefont {Leong}}, \bibinfo {author} {\bibfnamefont
  {T.}~\bibnamefont {Rudolph}}, \bibinfo {author} {\bibfnamefont
  {B.}~\bibnamefont {Eggleton}},\ and\ \bibinfo {author} {\bibfnamefont
  {C.}~\bibnamefont {Xiong}},\ }\bibfield  {title} {\bibinfo {title}
  {Indistinguishable heralded single photon generation via relative temporal
  multiplexing of two sources},\ }\href@noop {} {\bibfield  {journal} {\bibinfo
   {journal} {Optics express}\ }\textbf {\bibinfo {volume} {25}},\ \bibinfo
  {pages} {26067} (\bibinfo {year} {2017})}\BibitemShut {NoStop}%
\bibitem [{\citenamefont {Kaneda}\ and\ \citenamefont
  {Kwiat}(2019)}]{kaneda2019high}%
  \BibitemOpen
  \bibfield  {author} {\bibinfo {author} {\bibfnamefont {F.}~\bibnamefont
  {Kaneda}}\ and\ \bibinfo {author} {\bibfnamefont {P.~G.}\ \bibnamefont
  {Kwiat}},\ }\bibfield  {title} {\bibinfo {title} {High-efficiency
  single-photon generation via large-scale active time multiplexing},\
  }\href@noop {} {\bibfield  {journal} {\bibinfo  {journal} {Science Advances}\
  }\textbf {\bibinfo {volume} {5}},\ \bibinfo {pages} {eaaw8586} (\bibinfo
  {year} {2019})}\BibitemShut {NoStop}%
\bibitem [{\citenamefont {Adam}\ \emph {et~al.}(2014)\citenamefont {Adam},
  \citenamefont {Mechler}, \citenamefont {Santa},\ and\ \citenamefont
  {Koniorczyk}}]{adam2014optimization}%
  \BibitemOpen
  \bibfield  {author} {\bibinfo {author} {\bibfnamefont {P.}~\bibnamefont
  {Adam}}, \bibinfo {author} {\bibfnamefont {M.}~\bibnamefont {Mechler}},
  \bibinfo {author} {\bibfnamefont {I.}~\bibnamefont {Santa}},\ and\ \bibinfo
  {author} {\bibfnamefont {M.}~\bibnamefont {Koniorczyk}},\ }\bibfield  {title}
  {\bibinfo {title} {Optimization of periodic single-photon sources},\
  }\href@noop {} {\bibfield  {journal} {\bibinfo  {journal} {Physical Review
  A}\ }\textbf {\bibinfo {volume} {90}},\ \bibinfo {pages} {053834} (\bibinfo
  {year} {2014})}\BibitemShut {NoStop}%
\bibitem [{\citenamefont {Meyer-Scott}\ \emph {et~al.}(2020)\citenamefont
  {Meyer-Scott}, \citenamefont {Silberhorn},\ and\ \citenamefont
  {Migdall}}]{meyer2020single}%
  \BibitemOpen
  \bibfield  {author} {\bibinfo {author} {\bibfnamefont {E.}~\bibnamefont
  {Meyer-Scott}}, \bibinfo {author} {\bibfnamefont {C.}~\bibnamefont
  {Silberhorn}},\ and\ \bibinfo {author} {\bibfnamefont {A.}~\bibnamefont
  {Migdall}},\ }\bibfield  {title} {\bibinfo {title} {Single-photon sources:
  Approaching the ideal through multiplexing},\ }\href@noop {} {\bibfield
  {journal} {\bibinfo  {journal} {Review of Scientific Instruments}\ }\textbf
  {\bibinfo {volume} {91}},\ \bibinfo {pages} {041101} (\bibinfo {year}
  {2020})}\BibitemShut {NoStop}%
\bibitem [{\citenamefont {Brida}\ \emph {et~al.}(2010)\citenamefont {Brida},
  \citenamefont {Genovese},\ and\ \citenamefont
  {Berchera}}]{brida2010experimental}%
  \BibitemOpen
  \bibfield  {author} {\bibinfo {author} {\bibfnamefont {G.}~\bibnamefont
  {Brida}}, \bibinfo {author} {\bibfnamefont {M.}~\bibnamefont {Genovese}},\
  and\ \bibinfo {author} {\bibfnamefont {I.~R.}\ \bibnamefont {Berchera}},\
  }\bibfield  {title} {\bibinfo {title} {Experimental realization of
  sub-shot-noise quantum imaging},\ }\href@noop {} {\bibfield  {journal}
  {\bibinfo  {journal} {Nature Photonics}\ }\textbf {\bibinfo {volume} {4}},\
  \bibinfo {pages} {227} (\bibinfo {year} {2010})}\BibitemShut {NoStop}%
\bibitem [{\citenamefont {Samantaray}\ \emph {et~al.}(2017)\citenamefont
  {Samantaray}, \citenamefont {Ruo-Berchera}, \citenamefont {Meda},\ and\
  \citenamefont {Genovese}}]{samantaray2017realization}%
  \BibitemOpen
  \bibfield  {author} {\bibinfo {author} {\bibfnamefont {N.}~\bibnamefont
  {Samantaray}}, \bibinfo {author} {\bibfnamefont {I.}~\bibnamefont
  {Ruo-Berchera}}, \bibinfo {author} {\bibfnamefont {A.}~\bibnamefont {Meda}},\
  and\ \bibinfo {author} {\bibfnamefont {M.}~\bibnamefont {Genovese}},\
  }\bibfield  {title} {\bibinfo {title} {Realization of the first
  sub-shot-noise wide field microscope},\ }\href@noop {} {\bibfield  {journal}
  {\bibinfo  {journal} {Light: Science \& Applications}\ }\textbf {\bibinfo
  {volume} {6}},\ \bibinfo {pages} {e17005} (\bibinfo {year}
  {2017})}\BibitemShut {NoStop}%
\bibitem [{\citenamefont {Jakeman}\ and\ \citenamefont
  {Rarity}(1986)}]{jakeman1986use}%
  \BibitemOpen
  \bibfield  {author} {\bibinfo {author} {\bibfnamefont {E.}~\bibnamefont
  {Jakeman}}\ and\ \bibinfo {author} {\bibfnamefont {J.}~\bibnamefont
  {Rarity}},\ }\bibfield  {title} {\bibinfo {title} {The use of pair production
  processes to reduce quantum noise in transmission measurements},\ }\href@noop
  {} {\bibfield  {journal} {\bibinfo  {journal} {Optics communications}\
  }\textbf {\bibinfo {volume} {59}},\ \bibinfo {pages} {219} (\bibinfo {year}
  {1986})}\BibitemShut {NoStop}%
\bibitem [{\citenamefont {Sabines-Chesterking}\ \emph
  {et~al.}(2017)\citenamefont {Sabines-Chesterking}, \citenamefont {Whittaker},
  \citenamefont {Joshi}, \citenamefont {Birchall}, \citenamefont {Moreau},
  \citenamefont {McMillan}, \citenamefont {Cable}, \citenamefont {O’Brien},
  \citenamefont {Rarity},\ and\ \citenamefont {Matthews}}]{sabines2017sub}%
  \BibitemOpen
  \bibfield  {author} {\bibinfo {author} {\bibfnamefont {J.}~\bibnamefont
  {Sabines-Chesterking}}, \bibinfo {author} {\bibfnamefont {R.}~\bibnamefont
  {Whittaker}}, \bibinfo {author} {\bibfnamefont {S.}~\bibnamefont {Joshi}},
  \bibinfo {author} {\bibfnamefont {P.}~\bibnamefont {Birchall}}, \bibinfo
  {author} {\bibfnamefont {P.-A.}\ \bibnamefont {Moreau}}, \bibinfo {author}
  {\bibfnamefont {A.}~\bibnamefont {McMillan}}, \bibinfo {author}
  {\bibfnamefont {H.}~\bibnamefont {Cable}}, \bibinfo {author} {\bibfnamefont
  {J.}~\bibnamefont {O’Brien}}, \bibinfo {author} {\bibfnamefont
  {J.}~\bibnamefont {Rarity}},\ and\ \bibinfo {author} {\bibfnamefont
  {J.}~\bibnamefont {Matthews}},\ }\bibfield  {title} {\bibinfo {title}
  {Sub-shot-noise transmission measurement enabled by active feed-forward of
  heralded single photons},\ }\href@noop {} {\bibfield  {journal} {\bibinfo
  {journal} {Physical Review Applied}\ }\textbf {\bibinfo {volume} {8}},\
  \bibinfo {pages} {014016} (\bibinfo {year} {2017})}\BibitemShut {NoStop}%
\bibitem [{\citenamefont {Sabines-Chesterking}\ \emph
  {et~al.}(2019)\citenamefont {Sabines-Chesterking}, \citenamefont {McMillan},
  \citenamefont {Moreau}, \citenamefont {Joshi}, \citenamefont {Knauer},
  \citenamefont {Johnston}, \citenamefont {Rarity},\ and\ \citenamefont
  {Matthews}}]{sabines2019twin}%
  \BibitemOpen
  \bibfield  {author} {\bibinfo {author} {\bibfnamefont {J.}~\bibnamefont
  {Sabines-Chesterking}}, \bibinfo {author} {\bibfnamefont {A.}~\bibnamefont
  {McMillan}}, \bibinfo {author} {\bibfnamefont {P.}~\bibnamefont {Moreau}},
  \bibinfo {author} {\bibfnamefont {S.}~\bibnamefont {Joshi}}, \bibinfo
  {author} {\bibfnamefont {S.}~\bibnamefont {Knauer}}, \bibinfo {author}
  {\bibfnamefont {E.}~\bibnamefont {Johnston}}, \bibinfo {author}
  {\bibfnamefont {J.}~\bibnamefont {Rarity}},\ and\ \bibinfo {author}
  {\bibfnamefont {J.}~\bibnamefont {Matthews}},\ }\bibfield  {title} {\bibinfo
  {title} {Twin-beam sub-shot-noise raster-scanning microscope},\ }\href@noop
  {} {\bibfield  {journal} {\bibinfo  {journal} {Optics express}\ }\textbf
  {\bibinfo {volume} {27}},\ \bibinfo {pages} {30810} (\bibinfo {year}
  {2019})}\BibitemShut {NoStop}%
\bibitem [{\citenamefont {Moreau}\ \emph {et~al.}(2017)\citenamefont {Moreau},
  \citenamefont {Sabines-Chesterking}, \citenamefont {Whittaker}, \citenamefont
  {Joshi}, \citenamefont {Birchall}, \citenamefont {McMillan}, \citenamefont
  {Rarity},\ and\ \citenamefont {Matthews}}]{moreau2017demonstrating}%
  \BibitemOpen
  \bibfield  {author} {\bibinfo {author} {\bibfnamefont {P.-A.}\ \bibnamefont
  {Moreau}}, \bibinfo {author} {\bibfnamefont {J.}~\bibnamefont
  {Sabines-Chesterking}}, \bibinfo {author} {\bibfnamefont {R.}~\bibnamefont
  {Whittaker}}, \bibinfo {author} {\bibfnamefont {S.~K.}\ \bibnamefont
  {Joshi}}, \bibinfo {author} {\bibfnamefont {P.~M.}\ \bibnamefont {Birchall}},
  \bibinfo {author} {\bibfnamefont {A.}~\bibnamefont {McMillan}}, \bibinfo
  {author} {\bibfnamefont {J.~G.}\ \bibnamefont {Rarity}},\ and\ \bibinfo
  {author} {\bibfnamefont {J.~C.}\ \bibnamefont {Matthews}},\ }\bibfield
  {title} {\bibinfo {title} {Demonstrating an absolute quantum advantage in
  direct absorption measurement},\ }\href@noop {} {\bibfield  {journal}
  {\bibinfo  {journal} {Scientific reports}\ }\textbf {\bibinfo {volume} {7}},\
  \bibinfo {pages} {1} (\bibinfo {year} {2017})}\BibitemShut {NoStop}%
\bibitem [{\citenamefont {Losero}\ \emph {et~al.}(2018)\citenamefont {Losero},
  \citenamefont {Ruo-Berchera}, \citenamefont {Meda}, \citenamefont {Avella},\
  and\ \citenamefont {Genovese}}]{losero2018unbiased}%
  \BibitemOpen
  \bibfield  {author} {\bibinfo {author} {\bibfnamefont {E.}~\bibnamefont
  {Losero}}, \bibinfo {author} {\bibfnamefont {I.}~\bibnamefont
  {Ruo-Berchera}}, \bibinfo {author} {\bibfnamefont {A.}~\bibnamefont {Meda}},
  \bibinfo {author} {\bibfnamefont {A.}~\bibnamefont {Avella}},\ and\ \bibinfo
  {author} {\bibfnamefont {M.}~\bibnamefont {Genovese}},\ }\bibfield  {title}
  {\bibinfo {title} {Unbiased estimation of an optical loss at the ultimate
  quantum limit with twin-beams},\ }\href@noop {} {\bibfield  {journal}
  {\bibinfo  {journal} {Scientific reports}\ }\textbf {\bibinfo {volume} {8}},\
  \bibinfo {pages} {1} (\bibinfo {year} {2018})}\BibitemShut {NoStop}%
\bibitem [{\citenamefont {Magnoni}\ \emph {et~al.}(2019)\citenamefont
  {Magnoni}, \citenamefont {Grande}, \citenamefont {Knoll},\ and\ \citenamefont
  {Larotonda}}]{magnoni2019performance}%
  \BibitemOpen
  \bibfield  {author} {\bibinfo {author} {\bibfnamefont {A.~G.}\ \bibnamefont
  {Magnoni}}, \bibinfo {author} {\bibfnamefont {I.~H.~L.}\ \bibnamefont
  {Grande}}, \bibinfo {author} {\bibfnamefont {L.~T.}\ \bibnamefont {Knoll}},\
  and\ \bibinfo {author} {\bibfnamefont {M.~A.}\ \bibnamefont {Larotonda}},\
  }\bibfield  {title} {\bibinfo {title} {Performance of a temporally
  multiplexed single-photon source with imperfect devices},\ }\href@noop {}
  {\bibfield  {journal} {\bibinfo  {journal} {Quantum Information Processing}\
  }\textbf {\bibinfo {volume} {18}},\ \bibinfo {pages} {311} (\bibinfo {year}
  {2019})}\BibitemShut {NoStop}%
\bibitem [{\citenamefont {Achilles}\ \emph {et~al.}(2003)\citenamefont
  {Achilles}, \citenamefont {Silberhorn}, \citenamefont {{\'S}liwa},
  \citenamefont {Banaszek},\ and\ \citenamefont
  {Walmsley}}]{achilles2003fiber}%
  \BibitemOpen
  \bibfield  {author} {\bibinfo {author} {\bibfnamefont {D.}~\bibnamefont
  {Achilles}}, \bibinfo {author} {\bibfnamefont {C.}~\bibnamefont
  {Silberhorn}}, \bibinfo {author} {\bibfnamefont {C.}~\bibnamefont
  {{\'S}liwa}}, \bibinfo {author} {\bibfnamefont {K.}~\bibnamefont
  {Banaszek}},\ and\ \bibinfo {author} {\bibfnamefont {I.~A.}\ \bibnamefont
  {Walmsley}},\ }\bibfield  {title} {\bibinfo {title} {Fiber-assisted detection
  with photon number resolution},\ }\href@noop {} {\bibfield  {journal}
  {\bibinfo  {journal} {Optics letters}\ }\textbf {\bibinfo {volume} {28}},\
  \bibinfo {pages} {2387} (\bibinfo {year} {2003})}\BibitemShut {NoStop}%
\bibitem [{\citenamefont {Fitch}\ \emph {et~al.}(2003)\citenamefont {Fitch},
  \citenamefont {Jacobs}, \citenamefont {Pittman},\ and\ \citenamefont
  {Franson}}]{fitch2003photon}%
  \BibitemOpen
  \bibfield  {author} {\bibinfo {author} {\bibfnamefont {M.}~\bibnamefont
  {Fitch}}, \bibinfo {author} {\bibfnamefont {B.}~\bibnamefont {Jacobs}},
  \bibinfo {author} {\bibfnamefont {T.}~\bibnamefont {Pittman}},\ and\ \bibinfo
  {author} {\bibfnamefont {J.}~\bibnamefont {Franson}},\ }\bibfield  {title}
  {\bibinfo {title} {Photon-number resolution using time-multiplexed
  single-photon detectors},\ }\href@noop {} {\bibfield  {journal} {\bibinfo
  {journal} {Physical Review A}\ }\textbf {\bibinfo {volume} {68}},\ \bibinfo
  {pages} {043814} (\bibinfo {year} {2003})}\BibitemShut {NoStop}%
\bibitem [{\citenamefont {Lita}\ \emph {et~al.}(2008)\citenamefont {Lita},
  \citenamefont {Miller},\ and\ \citenamefont {Nam}}]{lita2008counting}%
  \BibitemOpen
  \bibfield  {author} {\bibinfo {author} {\bibfnamefont {A.~E.}\ \bibnamefont
  {Lita}}, \bibinfo {author} {\bibfnamefont {A.~J.}\ \bibnamefont {Miller}},\
  and\ \bibinfo {author} {\bibfnamefont {S.~W.}\ \bibnamefont {Nam}},\
  }\bibfield  {title} {\bibinfo {title} {Counting near-infrared single-photons
  with 95\% efficiency},\ }\href@noop {} {\bibfield  {journal} {\bibinfo
  {journal} {Optics express}\ }\textbf {\bibinfo {volume} {16}},\ \bibinfo
  {pages} {3032} (\bibinfo {year} {2008})}\BibitemShut {NoStop}%
\bibitem [{\citenamefont {Fukuda}\ \emph {et~al.}(2011)\citenamefont {Fukuda},
  \citenamefont {Fujii}, \citenamefont {Numata}, \citenamefont {Amemiya},
  \citenamefont {Yoshizawa}, \citenamefont {Tsuchida}, \citenamefont {Fujino},
  \citenamefont {Ishii}, \citenamefont {Itatani}, \citenamefont {Inoue} \emph
  {et~al.}}]{fukuda2011titanium}%
  \BibitemOpen
  \bibfield  {author} {\bibinfo {author} {\bibfnamefont {D.}~\bibnamefont
  {Fukuda}}, \bibinfo {author} {\bibfnamefont {G.}~\bibnamefont {Fujii}},
  \bibinfo {author} {\bibfnamefont {T.}~\bibnamefont {Numata}}, \bibinfo
  {author} {\bibfnamefont {K.}~\bibnamefont {Amemiya}}, \bibinfo {author}
  {\bibfnamefont {A.}~\bibnamefont {Yoshizawa}}, \bibinfo {author}
  {\bibfnamefont {H.}~\bibnamefont {Tsuchida}}, \bibinfo {author}
  {\bibfnamefont {H.}~\bibnamefont {Fujino}}, \bibinfo {author} {\bibfnamefont
  {H.}~\bibnamefont {Ishii}}, \bibinfo {author} {\bibfnamefont
  {T.}~\bibnamefont {Itatani}}, \bibinfo {author} {\bibfnamefont
  {S.}~\bibnamefont {Inoue}}, \emph {et~al.},\ }\bibfield  {title} {\bibinfo
  {title} {Titanium-based transition-edge photon number resolving detector with
  98\% detection efficiency with index-matched small-gap fiber coupling},\
  }\href@noop {} {\bibfield  {journal} {\bibinfo  {journal} {Optics express}\
  }\textbf {\bibinfo {volume} {19}},\ \bibinfo {pages} {870} (\bibinfo {year}
  {2011})}\BibitemShut {NoStop}%
\bibitem [{\citenamefont {Gerrits}\ \emph {et~al.}(2016)\citenamefont
  {Gerrits}, \citenamefont {Lita}, \citenamefont {Calkins},\ and\ \citenamefont
  {Nam}}]{gerrits2016superconducting}%
  \BibitemOpen
  \bibfield  {author} {\bibinfo {author} {\bibfnamefont {T.}~\bibnamefont
  {Gerrits}}, \bibinfo {author} {\bibfnamefont {A.}~\bibnamefont {Lita}},
  \bibinfo {author} {\bibfnamefont {B.}~\bibnamefont {Calkins}},\ and\ \bibinfo
  {author} {\bibfnamefont {S.~W.}\ \bibnamefont {Nam}},\ }\bibfield  {title}
  {\bibinfo {title} {Superconducting transition edge sensors for quantum
  optics},\ }in\ \href@noop {} {\emph {\bibinfo {booktitle} {Superconducting
  devices in quantum optics}}}\ (\bibinfo  {publisher} {Springer},\ \bibinfo
  {year} {2016})\ pp.\ \bibinfo {pages} {31--60}\BibitemShut {NoStop}%
\bibitem [{\citenamefont {Natarajan}\ \emph {et~al.}(2012)\citenamefont
  {Natarajan}, \citenamefont {Tanner},\ and\ \citenamefont
  {Hadfield}}]{natarajan2012superconducting}%
  \BibitemOpen
  \bibfield  {author} {\bibinfo {author} {\bibfnamefont {C.~M.}\ \bibnamefont
  {Natarajan}}, \bibinfo {author} {\bibfnamefont {M.~G.}\ \bibnamefont
  {Tanner}},\ and\ \bibinfo {author} {\bibfnamefont {R.~H.}\ \bibnamefont
  {Hadfield}},\ }\bibfield  {title} {\bibinfo {title} {Superconducting nanowire
  single-photon detectors: physics and applications},\ }\href@noop {}
  {\bibfield  {journal} {\bibinfo  {journal} {Superconductor science and
  technology}\ }\textbf {\bibinfo {volume} {25}},\ \bibinfo {pages} {063001}
  (\bibinfo {year} {2012})}\BibitemShut {NoStop}%
\bibitem [{\citenamefont {Marsili}\ \emph {et~al.}(2013)\citenamefont
  {Marsili}, \citenamefont {Verma}, \citenamefont {Stern}, \citenamefont
  {Harrington}, \citenamefont {Lita}, \citenamefont {Gerrits}, \citenamefont
  {Vayshenker}, \citenamefont {Baek}, \citenamefont {Shaw}, \citenamefont
  {Mirin} \emph {et~al.}}]{marsili2013detecting}%
  \BibitemOpen
  \bibfield  {author} {\bibinfo {author} {\bibfnamefont {F.}~\bibnamefont
  {Marsili}}, \bibinfo {author} {\bibfnamefont {V.~B.}\ \bibnamefont {Verma}},
  \bibinfo {author} {\bibfnamefont {J.~A.}\ \bibnamefont {Stern}}, \bibinfo
  {author} {\bibfnamefont {S.}~\bibnamefont {Harrington}}, \bibinfo {author}
  {\bibfnamefont {A.~E.}\ \bibnamefont {Lita}}, \bibinfo {author}
  {\bibfnamefont {T.}~\bibnamefont {Gerrits}}, \bibinfo {author} {\bibfnamefont
  {I.}~\bibnamefont {Vayshenker}}, \bibinfo {author} {\bibfnamefont
  {B.}~\bibnamefont {Baek}}, \bibinfo {author} {\bibfnamefont {M.~D.}\
  \bibnamefont {Shaw}}, \bibinfo {author} {\bibfnamefont {R.~P.}\ \bibnamefont
  {Mirin}}, \emph {et~al.},\ }\bibfield  {title} {\bibinfo {title} {Detecting
  single infrared photons with 93\% system efficiency},\ }\href@noop {}
  {\bibfield  {journal} {\bibinfo  {journal} {Nature Photonics}\ }\textbf
  {\bibinfo {volume} {7}},\ \bibinfo {pages} {210} (\bibinfo {year}
  {2013})}\BibitemShut {NoStop}%
\bibitem [{\citenamefont {Cahall}\ \emph
  {et~al.}(2017{\natexlab{a}})\citenamefont {Cahall}, \citenamefont {Nicolich},
  \citenamefont {Islam}, \citenamefont {Lafyatis}, \citenamefont {Miller},
  \citenamefont {Gauthier},\ and\ \citenamefont {Kim}}]{cahall2017multi}%
  \BibitemOpen
  \bibfield  {author} {\bibinfo {author} {\bibfnamefont {C.}~\bibnamefont
  {Cahall}}, \bibinfo {author} {\bibfnamefont {K.~L.}\ \bibnamefont
  {Nicolich}}, \bibinfo {author} {\bibfnamefont {N.~T.}\ \bibnamefont {Islam}},
  \bibinfo {author} {\bibfnamefont {G.~P.}\ \bibnamefont {Lafyatis}}, \bibinfo
  {author} {\bibfnamefont {A.~J.}\ \bibnamefont {Miller}}, \bibinfo {author}
  {\bibfnamefont {D.~J.}\ \bibnamefont {Gauthier}},\ and\ \bibinfo {author}
  {\bibfnamefont {J.}~\bibnamefont {Kim}},\ }\bibfield  {title} {\bibinfo
  {title} {Multi-photon detection using a conventional superconducting nanowire
  single-photon detector},\ }\href@noop {} {\bibfield  {journal} {\bibinfo
  {journal} {Optica}\ }\textbf {\bibinfo {volume} {4}},\ \bibinfo {pages}
  {1534} (\bibinfo {year} {2017}{\natexlab{a}})}\BibitemShut {NoStop}%
\bibitem [{\citenamefont {Tiffenberg}\ \emph {et~al.}(2017)\citenamefont
  {Tiffenberg}, \citenamefont {Sofo-Haro}, \citenamefont {Drlica-Wagner},
  \citenamefont {Essig}, \citenamefont {Guardincerri}, \citenamefont {Holland},
  \citenamefont {Volansky},\ and\ \citenamefont {Yu}}]{tiffenberg2017single}%
  \BibitemOpen
  \bibfield  {author} {\bibinfo {author} {\bibfnamefont {J.}~\bibnamefont
  {Tiffenberg}}, \bibinfo {author} {\bibfnamefont {M.}~\bibnamefont
  {Sofo-Haro}}, \bibinfo {author} {\bibfnamefont {A.}~\bibnamefont
  {Drlica-Wagner}}, \bibinfo {author} {\bibfnamefont {R.}~\bibnamefont
  {Essig}}, \bibinfo {author} {\bibfnamefont {Y.}~\bibnamefont {Guardincerri}},
  \bibinfo {author} {\bibfnamefont {S.}~\bibnamefont {Holland}}, \bibinfo
  {author} {\bibfnamefont {T.}~\bibnamefont {Volansky}},\ and\ \bibinfo
  {author} {\bibfnamefont {T.-T.}\ \bibnamefont {Yu}},\ }\bibfield  {title}
  {\bibinfo {title} {Single-electron and single-photon sensitivity with a
  silicon skipper ccd},\ }\href@noop {} {\bibfield  {journal} {\bibinfo
  {journal} {Physical Review Letters}\ }\textbf {\bibinfo {volume} {119}},\
  \bibinfo {pages} {131802} (\bibinfo {year} {2017})}\BibitemShut {NoStop}%
\bibitem [{\citenamefont {Hayat}\ \emph {et~al.}(1999)\citenamefont {Hayat},
  \citenamefont {Joobeur},\ and\ \citenamefont {Saleh}}]{hayat1999reduction}%
  \BibitemOpen
  \bibfield  {author} {\bibinfo {author} {\bibfnamefont {M.~M.}\ \bibnamefont
  {Hayat}}, \bibinfo {author} {\bibfnamefont {A.}~\bibnamefont {Joobeur}},\
  and\ \bibinfo {author} {\bibfnamefont {B.~E.}\ \bibnamefont {Saleh}},\
  }\bibfield  {title} {\bibinfo {title} {Reduction of quantum noise in
  transmittance estimation using photon-correlated beams},\ }\href@noop {}
  {\bibfield  {journal} {\bibinfo  {journal} {JOSA A}\ }\textbf {\bibinfo
  {volume} {16}},\ \bibinfo {pages} {348} (\bibinfo {year} {1999})}\BibitemShut
  {NoStop}%
\bibitem [{\citenamefont {Mood}\ \emph {et~al.}(1974)\citenamefont {Mood},
  \citenamefont {Graybill},\ and\ \citenamefont {Boes}}]{mood}%
  \BibitemOpen
  \bibfield  {author} {\bibinfo {author} {\bibfnamefont {A.}~\bibnamefont
  {Mood}}, \bibinfo {author} {\bibfnamefont {F.}~\bibnamefont {Graybill}},\
  and\ \bibinfo {author} {\bibfnamefont {D.}~\bibnamefont {Boes}},\ }\href@noop
  {} {\emph {\bibinfo {title} {Introduction to the Theory of Statistics}}}\
  (\bibinfo  {publisher} {McGraw-Hill},\ \bibinfo {year} {1974})\BibitemShut
  {NoStop}%
\bibitem [{\citenamefont {Frodesen}\ \emph {et~al.}(1979)\citenamefont
  {Frodesen}, \citenamefont {Skjeggestad},\ and\ \citenamefont
  {Tofte}}]{frodesen}%
  \BibitemOpen
  \bibfield  {author} {\bibinfo {author} {\bibfnamefont {A.~G.}\ \bibnamefont
  {Frodesen}}, \bibinfo {author} {\bibfnamefont {O.}~\bibnamefont
  {Skjeggestad}},\ and\ \bibinfo {author} {\bibfnamefont {H.}~\bibnamefont
  {Tofte}},\ }\href@noop {} {\emph {\bibinfo {title} {Probability and
  Statistics in Particle Physics}}}\ (\bibinfo  {publisher}
  {UNIVERSITETSFORLAGEN},\ \bibinfo {year} {1979})\ pp.\ \bibinfo {pages} {185
  -- 187}\BibitemShut {NoStop}%
\bibitem [{\citenamefont {Cahall}\ \emph
  {et~al.}(2017{\natexlab{b}})\citenamefont {Cahall}, \citenamefont {Nicolich},
  \citenamefont {Islam}, \citenamefont {Lafyatis}, \citenamefont {Miller},
  \citenamefont {Gauthier},\ and\ \citenamefont {Kim}}]{Cahall:17}%
  \BibitemOpen
  \bibfield  {author} {\bibinfo {author} {\bibfnamefont {C.}~\bibnamefont
  {Cahall}}, \bibinfo {author} {\bibfnamefont {K.~L.}\ \bibnamefont
  {Nicolich}}, \bibinfo {author} {\bibfnamefont {N.~T.}\ \bibnamefont {Islam}},
  \bibinfo {author} {\bibfnamefont {G.~P.}\ \bibnamefont {Lafyatis}}, \bibinfo
  {author} {\bibfnamefont {A.~J.}\ \bibnamefont {Miller}}, \bibinfo {author}
  {\bibfnamefont {D.~J.}\ \bibnamefont {Gauthier}},\ and\ \bibinfo {author}
  {\bibfnamefont {J.}~\bibnamefont {Kim}},\ }\bibfield  {title} {\bibinfo
  {title} {Multi-photon detection using a conventional superconducting nanowire
  single-photon detector},\ }\href {https://doi.org/10.1364/OPTICA.4.001534}
  {\bibfield  {journal} {\bibinfo  {journal} {Optica}\ }\textbf {\bibinfo
  {volume} {4}},\ \bibinfo {pages} {1534} (\bibinfo {year}
  {2017}{\natexlab{b}})}\BibitemShut {NoStop}%
\bibitem [{\citenamefont {Reddy}\ \emph {et~al.}(2019)\citenamefont {Reddy},
  \citenamefont {Lita}, \citenamefont {Nam}, \citenamefont {Mirin},\ and\
  \citenamefont {Verma}}]{Reddy:19}%
  \BibitemOpen
  \bibfield  {author} {\bibinfo {author} {\bibfnamefont {D.~V.}\ \bibnamefont
  {Reddy}}, \bibinfo {author} {\bibfnamefont {A.~E.}\ \bibnamefont {Lita}},
  \bibinfo {author} {\bibfnamefont {S.~W.}\ \bibnamefont {Nam}}, \bibinfo
  {author} {\bibfnamefont {R.~P.}\ \bibnamefont {Mirin}},\ and\ \bibinfo
  {author} {\bibfnamefont {V.~B.}\ \bibnamefont {Verma}},\ }\bibfield  {title}
  {\bibinfo {title} {Achieving 98\% system efficiency at 1550 nm in
  superconducting nanowire single photon detectors},\ }in\ \href
  {https://doi.org/10.1364/CQO.2019.W2B.2} {\emph {\bibinfo {booktitle}
  {Rochester Conference on Coherence and Quantum Optics (CQO-11)}}}\ (\bibinfo
  {publisher} {Optical Society of America},\ \bibinfo {year} {2019})\ p.\
  \bibinfo {pages} {W2B.2}\BibitemShut {NoStop}%
\bibitem [{\citenamefont {Blanchet}\ \emph {et~al.}(2008)\citenamefont
  {Blanchet}, \citenamefont {Devaux}, \citenamefont {Furfaro},\ and\
  \citenamefont {Lantz}}]{blanchet2008}%
  \BibitemOpen
  \bibfield  {author} {\bibinfo {author} {\bibfnamefont {J.-L.}\ \bibnamefont
  {Blanchet}}, \bibinfo {author} {\bibfnamefont {F.}~\bibnamefont {Devaux}},
  \bibinfo {author} {\bibfnamefont {L.}~\bibnamefont {Furfaro}},\ and\ \bibinfo
  {author} {\bibfnamefont {E.}~\bibnamefont {Lantz}},\ }\bibfield  {title}
  {\bibinfo {title} {Measurement of sub-shot-noise correlations of spatial
  fluctuations in the photon-counting regime},\ }\href
  {https://doi.org/10.1103/PhysRevLett.101.233604} {\bibfield  {journal}
  {\bibinfo  {journal} {Phys. Rev. Lett.}\ }\textbf {\bibinfo {volume} {101}},\
  \bibinfo {pages} {233604} (\bibinfo {year} {2008})}\BibitemShut {NoStop}%
\end{thebibliography}%

\end{document}